\documentclass[12pt]{article}\usepackage{amsmath,amsfonts}

\newcommand{\bse}{\begin{subequations}}
\newcommand{\ese}{\end{subequations}}
\newcommand{\be}{\begin{equation}}
\newcommand{\ee}{\end{equation}}
\newcommand{\bea}{\begin{eqnarray}}
\newcommand{\eea}{\end{eqnarray}}
\newcommand{\ba}{\begin{array}}
\newcommand{\ea}{\end{array}}
\newcommand{\bc}{\begin{center}}
\newcommand{\ec}{\end{center}}

\makeatletter \@addtoreset{equation}{section}

\makeatother
\def\by{\times}
\def\D{{\cal D}}
\def\F{{\cal F}}

\def\J{{\cal J}}
\def\K{{\cal K}}

\def\N{{\cal N}}
\def\Z{{\cal Z}}

\def\R{{\cal R}}
\def\P{{\cal P}}
\def\W{{\cal W}}

\def\Tr{{\rm Tr\ }}
\def\super{$PSU(2|2)\times PSU(2|2)\times U(1)$}
\def\ads5{$AdS_5\times S^5$\ }
\def\ie{{\it i.e.\ }}
\begin{document}
\baselineskip 18pt%

\begin{titlepage}
\vspace*{1mm}%
\hfill%
\vbox{
    \halign{#\hfil        \cr
           IPM/P-2006/034 \cr
           hep-th/0606117\cr
           }
      }
\vspace*{15mm}%
\begin{center}
{\Large{\bf Tiny Graviton Matrix Theory/SYM Correspondence:
\\ Analysis of BPS States}}%
\vspace*{10mm}%

{\bf M. Ali-Akbari$^{1,2}$, M. M. Sheikh-Jabbari$^{1}$, M. Torabian$^{1,2}$}%

\vspace*{0.4cm}
{\it {$^1$Institute for Studies in Theoretical Physics and Mathematics (IPM)\\
P.O.Box 19395-5531, Tehran, IRAN\\
$^2$Department of Physics, Sharif University of Technology\\
P.O.Box 11365-9161, Tehran, IRAN}}\\
{E-mails: {\tt aliakbari, jabbari, mahdi@theory.ipm.ac.ir}}%
\vspace*{1.5cm}
\end{center}

\begin{center}{\bf Abstract}\end{center}
\begin{quote}
In this paper we continue analysis of the Matrix theory describing
the DLCQ of type IIB string theory on $AdS_5\times S^5$ (and/or
the plane-wave) background, \ie the Tiny Graviton Matrix Theory
(TGMT) \cite{TGMT}. We study and classify  1/2, 1/4 and 1/8 BPS
solutions of the TGMT which are generically of the form of
rotating three brane giants. These are branes whose shape are
deformed three spheres and hyperboloids. In lack of a
classification of such ten dimensional type IIb supergravity
configurations, we focus on the dual $\N=4$ four dimensional 1/2,
1/4 and one 1/8 BPS operators and show that they are in one-to-one
correspondence with the states of the same set of quantum numbers
in TGMT. This provides further evidence in support of the Matrix
theory.
\end{quote}
\end{titlepage}
\tableofcontents
\section{Introduction}
The AdS/CFT \cite{AdS/CFT} has provided us with a very elegant
framework for studying quantum gravity (string theory) via a dual
gauge theory and vice-versa. This correspondence (duality),
however, is a kind of strong-weak duality in the sense that
generically only one side of the duality is perturbatively
accessible. Therefore, most of the analysis and checks of the
duality has been limited to the BPS sectors, the information of
which can be safely used at strongly coupled regime.

It has, however, been noted that in some specific limits/sectors
both sides of the AdS/CFT duality can be perturbative and hence
giving a window for a direct check of the duality. The most
extensively studied such sector is the so-called BMN sector
\cite{BMN}. The BMN sector is obtained through the translation of
taking the Penrose limit (which is an operation on the
geometry/gravity side) into the dual $\N=4$ SYM language (for
review e.g. see \cite{Review}).

Despite the large number of works devoted to the analysis of the
AdS/CFT duality, a non-perturbative description of either side is
still lacking. Inspired by the example of the BFSS matrix theory
\cite{BFSS}, according which the discrete light-cone quantization
(DLCQ) of M-theory is described by a $0+1$ SYM theory, one may
hope that type IIB string theory on $AdS_5\times S^5$, at least in
the DLCQ description, admits a Matrix theory formulation. In
\cite{TGMT} it was argued that this is indeed the case. To argue
for existence of such a Matrix theory description three
observations were noted in \cite{TGMT}: 1) The DLCQ description is
very similar to a description in infinite momentum frame (IMF)
\cite{BFSS} and in the IMF what is viewed from the  $AdS$ geometry
is its Penrose limit, the corresponding (maximally supersymmetric)
plane-wave geometry. 2) The DLCQ of M-theory on the maximally
supersymmetric {\it eleven dimensional} plane-wave geometry, the
geometry which is obtained as Penrose limit of either of the
$AdS_{4,7}\times S^{7,4}$ geometries, is described by a $0+1$
supersymmetric gauge theory, the BMN Matrix theory \cite{BMN}. 3)
It was shown in \cite{DSV1} that, similarly to the BFSS case
\cite{dHN}, the BMN matrix theory is a theory of discretized
membranes in the 11 dimensional plane-wave background. Moreover,
it was noted that the 1/2 BPS vacuum solutions of the BMN matrix
model are of the form of concentric giant membrane gravitons
\cite{DSV1}. These giant membrane gravitons in the BMN matrix
model notation appear as fuzzy two spheres. As pointed out in
\cite{TGMT} these membranes can be interpreted as a collection of
``tiny membrane gravitons'' blown up (by the Myers effect
\cite{Myers}) to cover a two sphere. Tiny membrane gravitons play
the role of  D0-branes in the BFSS theory. Hence, the BMN (or
plane-wave) matrix model, which is describing the DLCQ of M-theory
on the $AdS_{4,7}\times S^{7,4}$ and/or the 11 dimensional
plane-wave, is nothing but a {\it tiny (membrane) graviton matrix
theory}.

A similar idea may also be applied to obtain the DLCQ formulation
of type IIB string theory on the \ads5 background and/or the
corresponding plane-wave. In this matrix theory, however, we
should  employ the {\it tiny three-brane gravitons}. The action
for the tiny three-brane graviton theory, or the TGMT in short,
 in analogy with the 11 dimensional case, is obtained from
discretization of a three-brane action on the 10 dimensional
plane-wave background \cite{TGMT}. In section 2, we review the
statement of the TGMT conjecture, its action and symmetry
structure. As the DLCQ formulation of string theory on the
$AdS_5\times S^5$, one expects the TGMT to be in correspondence
with both type IIb supergravity and the dual $\N=4,\ D=4$ SYM
theory (see Fig.6 of \cite{half-BPS} which illustrates
AdS-TGMT-CFT triality).

In this paper we continue the analysis started in \cite{TGMT,
half-BPS} to provide further supportive evidence for the conjecture.
We study the BPS configurations of the TGMT. In section 3, we review
and expand results of \cite{half-BPS} by exhausting the 1/2 BPS
configurations of the TGMT and showing that they are very closely
related to the same configurations in the type IIb supergravity, the
LLM geometries \cite{LLM}, and the $\N=4$ dual gauge theory
\cite{Beren,Antal}. Here we show that the non-commutativity of the
$(x_1,x_2)$-plane in the LLM geometries \cite{x1x2-NC} comes as a
natural outcome in the tiny graviton matrix theory.

In section 4, we extend our analysis to the less BPS, i.e. 1/8 and
1/4 BPS, configurations. The BPS configurations that we study here
are all of the form of \emph{transverse} three branes of various
shape and topology and  fall into two classes, those which are
completely specified with the \emph{shape} of the brane, i.e.
geometric fluctuations of three branes. And those in which the
non-geometric, internal degrees of freedom of the three brane (\ie
the gauge fields on the D3-brane) is turned on. We give explicit
matrix representations of these deformed three sphere giants and
show that they are completely labelled by the number of giants and
at most four integers corresponding to their angular momenta.

In section 5, we review the BPS states/multiplets of the $\N=4$
SYM and show that all the BPS states we have studied in the TGMT
have correspondents in the dual gauge theory. This, via the
AdS/CFT, provides further supportive evidence for the TGMT
conjecture. We end this article by summary of our results and the
outlook.

In the TGMT Lagrangian {\it four brackets} has been introduced and
used. Since four brackets, unlike the usual two brackets, \ie
commutators, are not so familiar we find it useful to present some
identities regarding computations with these brackets. These and
some notation-fixings are gathered in Appendix A and Appendix B
contains the explicit form of the TGMT superalgebra, \super , in
terms of the matrices.

{\bf Note added:} When this work was finished \cite{Mandal}
appeared on the arxiv which has some overlap in the subject of the
current work.
\section{Review of the Tiny Graviton Matrix Theory}\label{review}%
In this section we briefly review  basics of the tiny graviton
Matrix theory, TGMT. It is essentially a very short summary of
\cite{TGMT}. The TGMT proposal states  that the DLCQ of type IIB
strings on the $AdS_5\times S^5$ or  the 10 dimensional plane-wave
background in the sector with $J$ units of light-cone momentum is
described by a $U(J)$ supersymmetric gauge theory.

\subsection{The TGMT action}
Dynamics of the TGMT is governed by the action which is the
regularized (discretized)  form of a \emph{single} D3-brane action
on the plane-wave background, once the light-cone gauge is fixed
and  while the gauge field living on the brane is turned off.
Fixing the light-cone gauge does not remove all the unphysical
gauge degrees of freedom. It fixes the part of four dimensional
diffeomorphisms which mixes temporal and spatial directions on the
brane world-volume. The parts which rotate spatial directions
among themselves remains unfixed. The spatial part of the
diffeomorphisms, after the prescribed ``regularization'' method
for discretizing the three-brane \cite{TGMT}, constitute the
$U(J)$ gauge symmetry of the TGMT. As is evident from the above
construction we expect in $J\to\infty$ limit to recover the
volume-preserving diffeomorphisms. This parallels the discussions
that the $U(N)$ gauge symmetry of the BFSS matrix model is nothing
but the discretized version of the area preserving diffeomorphism
on a membrane \cite{BFSS, dHN}. Again, in analogy with the BFSS
case, the TGMT can also be understood as a theory of $J$ tiny
gravitons. Tiny gravitons, similarly to the D0-branes, are on one
hand gravitons and on the other hand (tiny, spherical) D3-branes
and hence show the remarkable property of gauge symmetry
enhancement when become coincident. According to the TGMT
conjecture everything, including the fabric of space-time, is made
out of different configurations of tiny (three-brane) gravitons.

The TGMT action is then \be
{\cal S} = \int d\tau\ {\bf L},  \ee%
with the Lagrangian%
\be\label{Lagrangian}
\begin{split}
{\bf L} = R_-\ \Tr&\biggl[
\frac{1}{2R_-^2}\left[(\D_0X_i)^2+(\D_0X_a)^2\right] -
\frac{1}{2}\left(\frac{\mu}{R_-}\right)^2(X_i^2+X_a^2) \cr &-
\frac{1}{2\cdot 3!g_s^2} \left([ X^i , X^j , X^k, {\cal L}_5][ X^i
, X^j , X^k, {\cal L}_5] + [ X^a , X^b , X^c, {\cal L}_5][ X^a ,
X^b , X^c, {\cal L}_5]\right) \cr &- \frac{1}{2\cdot 2g_s^2}
\left([ X^i , X^j , X^a, {\cal L}_5][ X^i , X^j , X^a, {\cal L}_5]
+ [ X^a , X^b , X^i, {\cal L}_5][ X^a , X^b , X^i, {\cal
L}_5]\right) \cr & +\frac{\mu}{3!R_- g_s}\left( \epsilon^{i j k l}
X^i [X^j, X^k, X^l, {\cal L}_5]+ \epsilon^{a b c d} X^a [ X^b,
X^c, X^d , {\cal L}_5] \right)\cr & +
\left(\frac{i}{R_-}\right)\left(\theta^\dagger {}^{\alpha
\beta}\D_0 \theta_{\alpha \beta} + \theta^\dagger {}^{\dot\alpha
\dot\beta}\D_0 \theta_{\dot\alpha \dot\beta}\right) -
\left(\frac{\mu}{R_-}\right) \left(\theta^\dagger {}^{\alpha
\beta} \theta_{\alpha \beta}- \theta_{\dot\alpha
\dot\beta}\theta^\dagger {}^{\dot\alpha \dot\beta}\right)\cr
&-\frac{1}{2g_s}\left( \theta^\dagger {}^{\alpha \beta}
(\sigma^{ij})_\alpha^{\:  \: \delta} [ X^i, X^j, \theta_{\delta
\beta}, {\cal L}_5] + \theta^\dagger {}^{\alpha \beta}
(\sigma^{ab})_\alpha^{ \: \: \delta} \: [ X^a, X^b, \theta_{\delta
\beta}, {\cal L}_5]\right) \cr &- \frac{1}{2g_s}
\left(\theta^\dagger {}^{\dot\alpha \dot\beta}
(\sigma^{ij})_{\dot\alpha}^{ \: \: \dot\delta} \: [ X^i, X^j,
\theta_{\dot\delta \dot\beta}, {\cal L}_5]+ \theta^\dagger
{}^{\dot\alpha \dot\beta} (\sigma^{ab})_{\dot\alpha}^{\: \:
\dot\delta} \: [ X^a, X^b, \theta_{\dot\delta \dot\beta},
{\cal L}_5]\right)\biggr]\ ,\end{split}\ee%
where the $X$'s and $\theta$'s are $J\by J$ matrices in the
adjoint of $U(J)$, $i=1,2,3,4$ and $a=5,6,7,8$. The above action
besides the $U(J)$ gauge symmetry has $SO(4)_i\times SO(4)_a$
global symmetry, as well as a $\mathbb{Z}_2$ which exchanges $i,a$
directions and the fermionic indices $\alpha, \beta$ which run
over $1,2$ are Weyl indices of either of the $SO(4)_i$ or
$SO(4)_a$ symmetries.

In our conventions $R_-$, which is the radius of compactification
of the light-like direction $X^-$ in the plane-wave geometry,
similarly to the $\mu$, has dimension of energy. In fact the $X^-$
compactification radius (in string units) is $R_-/\mu$ which is a
free parameter once TGMT is used as string theory on  the
plane-wave background. Besides $\mu/R_-$ in the action we have
another dimensionless parameter, $g_s$. As discussed in
\cite{TGMT} and reviewed in the introduction, TGMT may also be
used as DLCQ formulation of strings on the $AdS_5\times S^5$
geometry. In this case $R_-/\mu$ is
related to the $AdS$ radius, $(l_s^4g_s N)^{1/4}$, as \cite{TGMT}%
\be\label{R-N}%
\left(\frac{R_-}{\mu}\right)^2=g_s N=\frac{R^4_{AdS}}{l_s^4}\ .%
\ee%

In the above action ${\cal L}_5$ is a fixed (non-dynamical)
$SO(4)_i\times SO(4)_a$ invariant, Hermitian $J\times J$ matrix
defined through \cite{TGMT,half-BPS}\footnote{These relations are
$U(J)$ invariant and hence one can choose the basis in which
${\cal L}_5$ is diagonal. This specifies  ${\cal L}_5$ up to
permutation of the eigenvalues, ${\cal S}_J$.}$^,$\footnote{There
has been another proposal for the plane-wave matrix string theory
which does not involve the ${\cal L}_5$ \cite{Lozano-1}. This
theory is governed by the action for $J$ coincident ``gravitons''
\cite{Lozano-2}. Matrix model of \cite{Lozano-1},
however, only exhibits $(SU(2)\times U(1) )^2$ out of $SO(4)\times SO(4)$.}%
\be\label{L5}%
\Tr{\cal L}_5^2= {1}\ ,\ \ \ \ \Tr{\cal L}_5=0.
\ee%
Also, $\D_0$ is the covariant derivative of the  $0+1$ dimensional
$U(J)$ gauge theory%
\be \D_0 = \partial_0 + i[{\cal A}_0,\ .\ ] \ee%
${\cal A}_0 = {\cal A}_0^m T_m$ is the only component of the gauge
field and $T_m,$ $m=1,2,\dots,J^2$ are the generators of the
$U(J)$ gauge symmetry. We can use the temporal gauge and set
${\cal A}_0 = 0$.

To ensure this gauge condition, all of our physical states must
satisfy the Gauss law constraint arising from equations of motion
of ${\cal A}_0$. These constraints, which consist of $J^2-1$
independent conditions are%
\be\label{Gauss-law}%
\left(\Phi_{J\by J} \equiv
i[X^i,P^i]+i[X^a,P^a]+\{\theta^{\dagger\alpha\beta},\theta_{\alpha\beta}\}
+\{\theta^{\dagger\dot\alpha\dot\beta},\theta_{\dot\alpha\dot\beta}\}\right)
|Phys\rangle
= 0 \ee%
where $P^I = D_0X^I$. Note that \eqref{Gauss-law} is an operator
equation which should be satisfied with all physical states and
the commutators are matrix commutators.

It is useful to derive the light-cone Hamiltonian of
the theory using  Legender transformation%
\be%
{\bf H} = \Tr P^I\dot X^I- {\bf L} + \Tr{\cal A}_0\Phi%
\ee%
where $I=1,2,\cdots 8$,
\[
P_I=\frac{\partial {\bf L}}{\partial \dot X^I}
\]
and the last term is Lagrange multiplier ${\cal A}_0$, times the
equation of motion of ${\cal A}_0$. The explicit form of the
Hamiltonian in the temporal gauge is given in the Appendix
\ref{AppendixB}.

\subsection{The symmetry structure}
The plane-wave is a maximally supersymmetric background, \ie it
has 32 fermionic isometries, which can be arranged into two sets
of 16 in kinematical supercharges and dynamical supercharges, and
30 bosonic isometries. More details can be found in \cite{Review}.

On the other hand, TGMT, has a large number of local and global
symmetries. Let us work with the Hamiltonian in ${\cal A}_0=0$
gauge given in \eqref{Hamiltonian}. This Hamiltonian still enjoys
the {\it time independent} part of the $U(J)$ gauge symmetry,
which appears as a global symmetry. Moreover, it  has
$PSU(2|2)\times PSU(2|2)\times U(1)_{\bf H}\times U(1)_{p^+}$,
with the generators $Q_{\alpha\dot\beta},\
Q_{\dot\alpha\beta},\ J^{ij},\ J^{ab},\ {\bf H},\ p^+$. That is,%
 \be
[P^{+}, Q_{\alpha\dot\beta}]=0,\quad [P^{+},
Q_{\dot\alpha\beta}]=0,\quad [{\bf H},
Q_{\alpha\dot\beta}]=0,\quad
[{\bf H},Q_{\dot\alpha\beta}]=0 .\ee%
\bse\label{superalgebra}\begin{align}
\label{Q1}\{Q_{\alpha\dot\beta},Q^{\dagger\rho\dot\lambda}\} &=
\delta_{\alpha}^{\rho} \delta_{\dot\beta}^{\dot\lambda}{\bf H} +
\frac{\mu}{2}(i\sigma^{ij})_{\alpha}^{\rho}
\delta_{\dot\beta}^{\dot\lambda}{\bf J}_{ij} +
\frac{\mu}{2}\delta_{\alpha}^{\rho}
(i\sigma^{ab})_{\dot\beta}^{\dot\lambda}{\bf J}_{ab} -
{\mu}(i\sigma^{ij})_{\alpha}^{\rho}
(i\sigma^{ab})_{\dot\beta}^{\dot\lambda}\R_{ijab} \\
\label{Q2}\{Q_{\dot{\alpha}\beta},Q^{\dagger\dot{\rho}\lambda}\}
&= \delta_{\dot{\alpha}}^{\dot{\rho}} \delta_{\beta}^{\lambda}{\bf
H} + \frac{\mu}{2}(i\sigma^{ij})_{\dot{\alpha}}^{\dot{\rho}}
\delta_{\beta}^{\lambda}{\bf J}_{ij} +
\frac{\mu}{2}\delta_{\dot{\alpha}}^{\dot{\rho}}
(i\sigma^{ab})_{\beta}^{\lambda}{\bf J}_{ab} -
{\mu}(i\sigma^{ij})_{\dot{\alpha}}^{\dot{\rho}}
(i\sigma^{ab})_{\beta}^{\lambda}\R_{ijab} \\
\{Q_{\alpha\dot\beta},Q^{\dagger\dot\rho\lambda}\} &= \mu
(\sigma^{i})_{\alpha}^{\dot\rho}(\sigma^{a})_{\dot\beta}^{\lambda}
(C_{ia}+i\hat{C}_{ia}) \\
\{Q_{\dot{\alpha}\beta},Q^{\dagger\rho\dot{\lambda}}\} &= \mu
(\sigma^{i})_{\dot{\alpha}}^{\rho}(\sigma^{a})_{\beta}^{\dot{\lambda}}
(C_{ia}-i\hat{C}_{ia}) \\
\{Q_{\dot{\alpha}\beta},Q_{\dot{\rho}\lambda}\} &= 0,\quad
\{Q_{{\alpha}\dot\beta},Q_{{\rho}\dot\lambda}\} = 0,\quad
\{Q_{\dot{\alpha}\beta},Q_{{\rho}\dot \lambda}\} = 0 .\end{align}\ese%
The other bosonic generators, $R_{ijab}, C_{ia}, \hat{C}_{ia}$ are
constituting some of the possible  extensions of the minimal
\super\ which are inherent to the TGMT formulation
\cite{extensions}. In our conventions the complex conjugate
$\dagger$ just raises and lowers the fermionic indices. 
{To find the explicit form of (some of) the extensions in terms of
matrices \ one should perform a careful computation of various
anticommutators of the above dynamical supercharges, this has been
carried out in \cite{extensions}. (A similar calculation can be
carried out for kinematical and mixed supercharges.) For the
computation we need to use the following basic \emph{operator} (to
be compared
with matrix) commutation relations:%
\bea\label{basic-commutators}%
[X^I_{pq}, P^J_{rs}] &=& i\delta^{IJ}\ \delta_{ps}\delta_{qr} \cr
\{(\theta^{\dagger \alpha\beta})_{pq},
(\theta_{\rho\gamma})_{rs}\} &=& \delta^{\alpha}_{\rho}
\delta^{\beta}_{\gamma}\ \delta_{ps}\delta_{qr} \\
\{(\theta^{\dagger \dot\alpha\dot\beta})_{pq},
(\theta_{\dot\rho\dot\gamma})_{rs}\} &=&
\delta^{\dot\alpha}_{\dot\rho} \delta^{\dot\beta}_{\dot\gamma}\
\delta_{ps}\delta_{qr},\ \ \ p,r,s,t=1,2,\cdots J\nonumber
\eea%

The two $SO(4)_i$ and $SO(4)_a$ rotations act on $i$ and $a$
vector indices of the bosonic $X^i, P_i$ and $X^a, P_a$ fields and
on the spinor (Weyl) indices of fermionic $\theta_{\alpha\beta}$
as%
\be\label{SO4-rotation}%
\begin{split}
X^i_{rs}&\to \tilde X^i_{rs}= R^i_j X^j_{rs}\\
(\theta_{\alpha\beta})_{rs}&\to
(\tilde\theta_{\alpha\beta})_{rs}=R_{\alpha\gamma} (\theta_{\gamma\beta})_{rs}\\
 {\cal L}_5&\to {\cal L}_5 ,
\end{split}
\ee%
where $R_{ij}=e^{i\omega_{ij} \gamma^{ij}}, \
R_{\alpha\gamma}=e^{i\omega_{ij} \sigma^{ij}}$ are respectively
$4\by 4$ and $2\by 2$ $SO(4)$ rotation matrices and $r,s$ are
$J\by J$ indices. There is another $\mathbb{Z}_2$ symmetry which
changes the orientation of the $X^i$ and $X^a$
\emph{simultaneously} (\ie $\epsilon_{ijkl},\ \epsilon_{abcd}\to
-\epsilon_{ijkl},\ -\epsilon_{abcd})$ together with sending ${\cal
L}_5\to -{\cal L}_5$.

Under the $U(J)$ rotations all the dynamical fields as well as the
${\cal L}_5$ are in the
adjoint representation:%
\be\label{U(J)-rotation}%
X_I\to \hat X^I=U X^I U^{-1},\ \ \ {\cal L}_5\to U {\cal L}_5 U^{-1}%
\ee%
where $U\in U(J)$. There is a $U(1)$ subgroup of $U(J)$,
$U(1)_\alpha$
which is generated by ${\cal L}_5$:%
\be\label{Ualpha}
U_\alpha = e^{i\alpha{\cal L}_5}.%
\ee%
This subgroup has the interesting property that keeps the ${\cal
L}_5$ invariant. We will discuss the $U(1)_\alpha$ further in the
following sections.

\subsection{Classical BPS equations: Matrix regularized
form}\label{section2.3}%
Given the superalgebra we are now ready to study BPS
configurations of the TGMT, which is the main theme of this paper.
By definition, a BPS state is a field configuration which is
invariant under some specific supersymmetry transformations. For
the configurations in which the spinors $\theta$'s are turned off
the non-zero SUSY variations are only $\delta_\epsilon\theta$'s
and hence the BPS equations read as%
\be\label{Q-BPS}%
\delta_\epsilon\theta = \left\{\epsilon^\dagger Q + \epsilon
Q^\dagger,\theta\right\}=0%
\ee%
for classical configurations, and $\delta_\epsilon\theta
|BPS\rangle=0$ for quantum BPS states, for some spinor $\epsilon$.
The number of independent $\epsilon$'s which satisfy either of the
above equations determines how much BPS our configuration is.
Explicitly, if there are $n$ $\epsilon$'s the configuration is
$n/32$ BPS.

TGMT is a DLCQ formulation of strings and as such the kinematical
supercharges $q_{\alpha\beta}, q^{\dagger\alpha\beta},
q_{\dot\alpha\dot\beta}, q^{\dagger\dot\alpha\dot\beta}$ which
anticommute to light-cone momentum $p^+$ are not preserved by any
physical state. Hence the BPS configurations of TGMT are 1/2 or
less BPS. To check what portion of the dynamical supercharges,
$Q$'s, is preserved we write out \eqref{Q-BPS} explicitly
\be\label{variation1}\begin{split}
\delta_\epsilon(\theta_{\alpha\beta})_{pq} &=
\left\{\epsilon^{\dagger\gamma\dot\delta}Q_{\gamma\dot\delta} +
\epsilon^{\dagger\dot\gamma\delta}Q_{\dot\gamma\delta} +
\epsilon_{\gamma\dot\delta}Q^{\dagger\gamma\dot\delta} +
\epsilon_{\dot\gamma\delta}Q^{\dagger\dot\gamma\delta} ,
(\theta_{\alpha\beta})_{pq}\right\} \cr
&=\Big((P^d+i\frac{\mu}{R_-}X^d) +
\frac{i}{3!g_s}\epsilon^{abcd}[X^a,X^b,X^c,{\cal L}_5]\Big)_{pq}
(\sigma^d)^{\dot\delta}_\beta\epsilon_{\alpha\dot\delta} \cr &+
\frac{1}{2g_s}[X^a,X^i,X^j,{\cal L}_5]_{pq}
(\sigma^a)^{\dot\delta}_\beta(i\sigma^{ij})^\gamma_\alpha\epsilon_{\gamma\dot\delta}
\cr &+ \Big((P^l+i\frac{\mu}{R_-}X^l) +
\frac{i}{3!g_s}\epsilon^{ijkl}[X^i,X^j,X^k,{\cal L}_5]\Big)_{pq}
(\sigma^l)^{\dot\gamma}_\alpha\epsilon_{\dot\gamma\beta} \cr &+
\frac{1}{2g_s}[X^i,X^a,X^b,{\cal L}_5]_{pq}
(\sigma^i)^{\dot\gamma}_\alpha(i\sigma^{ab})^\delta_\beta\epsilon_{\dot\gamma\delta}
\end{split}\ee
and%
\be\label{variation2}\begin{split}
\delta_\epsilon(\theta_{\dot\alpha\dot\beta})_{pq} &=
\left\{\epsilon^{\dagger\gamma\dot\delta}Q_{\gamma\dot\delta} +
\epsilon^{\dagger\dot\gamma\delta}Q_{\dot\gamma\delta} +
\epsilon_{\gamma\dot\delta}Q^{\dagger\gamma\dot\delta} +
\epsilon_{\dot\gamma\delta}Q^{\dagger\dot\gamma\delta} ,
(\theta_{\dot\alpha\dot\beta})_{pq}\right\} \cr
&=\Big((P^d+i\frac{\mu}{R_-}X^d) +
\frac{i}{3!g_s}\epsilon^{abcd}[X^a,X^b,X^c,{\cal L}_5]\Big)_{pq}
(\sigma^d)^\delta_{\dot\beta}\epsilon_{\dot\alpha\delta} \cr &+
\frac{1}{2g_s}[X^a,X^i,X^j,{\cal L}_5]_{pq}
(\sigma^a)^\delta_{\dot\beta}(i\sigma^{ij})^{\dot\gamma}_{\dot\alpha}
\epsilon_{\dot\gamma\delta} \cr &+ \Big((P^l+i\frac{\mu}{R_-}X^l)
+ \frac{i}{3!g_s}\epsilon^{ijkl}[X^i,X^j,X^k,{\cal L}_5]\Big)_{pq}
(\sigma^l)^\gamma_{\dot\alpha}\epsilon_{\gamma\dot\beta} \cr &+
\frac{1}{2g_s}[X^i,X^a,X^b,{\cal L}_5]_{pq}
(\sigma^i)^\gamma_{\dot\alpha}(i\sigma^{ab})^{\dot\delta}_{\dot\beta}
\epsilon_{\gamma\dot\delta}\ . \end{split}\ee%
The BPS equation, that is the demand that \eqref{variation1},
\eqref{variation2} vanish, will specify a set of $\epsilon$'s for
a given configuration.

To analyze and solve BPS equations, here we  use the method
introduced and used in \cite{Bak, Park}. The idea is that one can
always define a Hermitian projector $\P$ which is a $16\by 16$
matrix, and has $16-n$ zero eigenvalues for an $n/32$ BPS
configuration, that is $\P^2=\P,\ \Tr\P=n$. Choosing $\epsilon$ to
be of the form
\[
\epsilon=\P \epsilon_0,
\]
we demand that the BPS equation should then hold for any arbitrary
spinor $\epsilon_0$.
That is,%
\be\label{projectedBPS}%
\delta_\epsilon\theta = \cdots\P\epsilon_0 = 0
\ee%
where $\cdots$ stands for the expression in terms of $X$'s and
$P$'s given above. Since $\epsilon_0$ is an arbitrary spinor,
 drops out and hence the BPS equation  becomes an
equation only among  matrix degrees of freedom
\be\label{functional-equation}%
\F_{\P}[X,\theta,{\cal A}_0;{\cal L}_5] = 0.
\ee%
We should still make sure that the configuration which solves
\eqref{functional-equation} is physical in the sense that it is
compatible with the Gauss law%
\be\label{bosonic-Gauss-law}%
[X^i,\dot X^i+i[{\cal A}_0,X^i]]+[X^a,\dot X^a+i[{\cal A}_0,X^a]] = 0. %
\ee%
As it was shown in \cite{Bak,Park} the BPS equation
\eqref{functional-equation} can be considerably simplified by
choosing an appropriate form for ${\cal A}_0$ using the $U(J)$
gauge symmetry. It turns out (it will become clear
in section \ref{Less-BPS-section}) that%
\be\label{gauge-fixing}%
i[{\cal A}_0,\Psi]= -\frac{1}{g_s}[\Psi,X^1,X^2,{\cal
L}_5]-\frac{1}{g_s}[\Psi,X^3,X^4,{\cal L}_5],
\ee%
where $\Psi$ is a generic function of matrix degrees of freedom,
is a suitable choice for our purpose. In this gauge our BPS
equations reduces to a simple first order equations in time which
could be solved by a simple exponential.
 It, however, remains to ensure the Gauss law to
obtain the BPS configurations. This Gauss law is the main equation
to be solved and for the configurations in which $X^a$'s are
turned off, in the above gauge it takes the form:
\be\label{Gauss-law-BPS-gauge}%
[X^i,\dot X^i]-\frac{1}{g_s}\left([X^i,[X^i,X^1,X^2,{\cal
L}_5]]+[X^i,[X^i,X^3,X^4,{\cal L}_5]]\right)=0.
\ee%

Before discussing various solutions of the BPS equations, it is
worth noting that  \eqref{gauge-fixing} does not completely fix
the $U(J)$ gauge symmetry. As $[{\cal L}_5, X, Y,{\cal L}_5]=0$,
it fixes ${\cal A}_0$ only up to $U(1)_\alpha$ rotations ({\it
cf.} \eqref{Ualpha}).
\section{Half BPS Solutions}\label{half-BPS-section}%
This part is mainly a review of \cite{half-BPS}, we repeat it here
for completeness. We'll also add some more physical discussions
regarding the connection of these solutions and the LLM geometries
\cite{LLM}. Half BPS configurations are those which preserve all
the dynamical supercharges and hence both equations
\eqref{variation1} and \eqref{variation2} must vanish for these
configurations. In terms of the projector method, we should choose
$\P = {\bf 1}$ as the projection matrix. The BPS equations imply
that $X$'s should be time independent and
\bse\label{master-1/2BPS-equations}\begin{align}
[X^i,X^j,X^k,{\cal L}_5] & = -\frac{\mu g_s}{R_-}\ \epsilon^{ijkl} X^l \\
[X^a,X^b,X^c,{\cal L}_5]  & = -\frac{\mu g_s}{R_-}\ \epsilon^{abcd} X^d \\
[X^a,X^b,X^i,{\cal L}_5]  &= [X^a, X^i, X^j, {\cal L}_5] =0\ .
\end{align}\ese%
For half BPS configurations $P_i=P_a=0$ and hence  the Gauss law
is trivially satisfied.

The simplest solution is of course $X^i=X^a=0$. The first
non-trivial set of solutions is obtained by setting $X^a$ (or
$X^i$'s) equal to zero. In this case,
(\ref{master-1/2BPS-equations}b,c) is evident. To
solve (\ref{master-1/2BPS-equations}a), let us introduce%
\be\label{XvsJ}%
X^i=\Big(\frac{\mu g_s}{R_-}\Big)^{1/2}\J^j\ ,\qquad{\cal L}_5\equiv\J_5 ,%
\ee%
where $\J^i$'s are in $J\by J$ representations of $Spin(4)$ and
$\J_5$ is the chirality matrix,
 satisfying \cite{half-BPS} (see Appendix A for the explicit form of $\J^i$ matrices)%
\be\label{J-def}%
[\J^i,\J^j,\J^k,\J_5] = -\epsilon^{ijkl}\J^l .%
\ee%
The above specifies $\J$'s, they  can however be in reducible or
irreducible representations of $Spin(4)$. If in  the irreducible
representation, $\J$'s fulfill%
\be\label{J-irrep}%
\sum_{i=1}^4 \delta_{ij}\J^i\J^j = J\ {\bf 1}_{J\by J}.%
\ee%
In the case of reducible representations, $\sum_{i=1}^4
\delta_{ij}\J^i\J^j$ becomes block diagonal and in each block its
value is equal to the size of the block \cite{half-BPS}.

In the irreducible representation $X^i=\big(\frac{\mu
g_s}{R_-}\big)^{1/2}\J^j$ defines a \emph{single} {\it fuzzy three
sphere} of radius \footnote{It is remarkable that  in terms of the
\ads5 parameters, the ``fuzziness'' or the \emph{tiny graviton
scale} $l$ in Planck units is related to $1/N$ as \cite{TGMT}
({\it cf.} \eqref{R-N}) \[l^4=\frac{1}{N}\ l_p^4.\]
}$^,$\footnote{We should point out that, the appearance of ${\cal
L}_5$ in the definition of the fuzzy three sphere, as explained in
\cite{TGMT,half-BPS}, is related to the fact that in the
$SO(even)$ groups (in particular $SO(4)$) we have the chirality
operator. More intuitively, a fuzzy three sphere may be obtained
from a fuzzy four sphere once we restrict ourselves to a narrow
strip around the equator; due to the fuzziness this cutting cannot
be done exactly at the equator \cite{half-BPS}. At the level of
TGMT and quantum type IIB string theory, however,  ${\cal L}_5$ is
the reminiscent of the 11$^{th}$ circle once we compactify
M-theory on a shrinking two torus to obtain type IIB string theory
\cite{half-BPS}. In this viewpoint the spherical three branes
are M-theory five branes wrapping the $T^2$.}%
\be\label{radius-J}%
R^2=l^2\ J,\qquad\quad l^2\equiv \frac{\mu g_s}{R_-}\ l_s^2 .%
\ee%

In the reducible cases we have concentric multi fuzzy three sphere
configurations whose radii squared sum to $R^2$. These fuzzy
spheres are   in fact ``quantized'' spherical three brane
giants.\footnote{In terms of the matrix model presented in
\cite{Lozano-1} the vacuum solutions differ from our fuzzy three
spheres in the sense that they are of the form of three spheres
which are realized as fuzzy two spheres with an {\it Abelian}
$U(1)$ fiber over it.} The multi giant solutions is completely
specified once we give the
set of $\{J_k\}$'s whose sum is $J$:%
\be%
\sum_{k=1}\ J_k=J .%
\ee%
The above equation, the problem of partition of a given integer
into non-negative integers, is solved by Young tableaux of $J$
boxes, with $J_k$ number of boxes in each column (or row). Our
solution has a $\mathbb{Z}_2$ symmetry which exchanges the columns
and rows on the Young tableau, corresponding to giants and dual
giants exchange symmetry \cite{x1x2-NC, Beren2}.

One can now find the most general solutions with both $X^i$ and
$X^a$ non-zero, using $\J$'s. It is enough to take both of $X^i$
and $X^a$ to be proportional to $\J$'s but choose  $\J$'s to be in
\emph{reducible} representations in such a way that non-zero parts
of $X^i$'s and $X^a$'s do not overlap, \ie\ $[X^i,X^a]=0$. In this
case we have a collection of both giants and dual giants of
various radii. The giants are centered at $X^a=0$ while the dual
giants at $X^i=0$.

It should be noted that equations \eqref{XvsJ} and \eqref{J-def}
define the half BPS configurations up to $U(J)$ and $SO(4)$
rotations, e.g. obviously if $X^i$ is a fuzzy sphere solution
$\tilde X^i = UX^i U^{-1},\ U\in U(J)$ and $\tilde X^i=R_{ij}
X^j,\ R\in SO(4)$ also describe the same physical configuration.

In \cite{half-BPS} a one-to-one connection between half BPS
configurations of the TGMT and the LLM geometries which are
deformations about the plane-wave geometry were demonstrated.
These deformations in the LLM language are given by a ladder type
black and white configuration in the $(x_1,x_2)$-plane \cite{LLM}
and $x_1$ is to be identified with the $X^-$ direction. In the
TGMT $X^-$ is compactified on a circle of radius $R_-/\mu$ in
string units. The $x_2$ direction, however, is half of the
difference of the radii squared of the two three spheres in the
LLM geometry ({\it cf.} equation (2.20) of \cite{LLM}). Since we
should be able to wrap  our fuzzy spheres on these three spheres,
or equivalently for the LLM geometries viewed and probed by the
``quantized'' spherical three brane probes, equation
\eqref{radius-J} implies that the spectrum of $x_2$ should be
quantized in units of $l^2$. (Note that in the LLM conventions
$x_1$ and $x_2$ both have dimension length squared.) That
is%
\be\label{x1-quantized}%
x_2=l^2\times k,\qquad \quad k\in \mathbb{Z}.%
\ee%
In other words, $\Delta x_2=l^2$. Since $\Delta x_1=\frac{R_-}{\mu} l_s^2$,%
\be%
{\rm smallest\ volume\ on\ }(x_1,x_2){\rm -plane}=\Delta x_1\Delta
x_2= l_s^4 g_s=l_p^4\ . %
\ee%
This is remarkable because that is exactly the result coming from
the  semiclassical analysis of quantization of the (five-form)
flux in the LLM geometries \cite{LLM}. This result can, however,
 also come from  $[x_1,x_2]=il_p^4$ relation, if one assumes
$(x_1,x_2)$-plane to be a noncommutative Moyal plane. In the case
of our interest we are in fact dealing with a noncommutative
cylinder (e.g. see \cite{Classify}) and the noncommutative
structure is a direct outcome of the TGMT formulation.

In the TGMT setup with finite $J$ there is an upper limit on the
radii of the three spheres set by $J$. That is, to cover the whole
LLM geometry one should take infinite $J$ limit. This limit could
be taken either keeping $R_-/\mu$ fixed or  sending it to
infinity, keeping $g_s$ fixed.

The other conclusion inferred from the quantization of the three
sphere radii in the LLM geometries is that the spectrum of the $y$
coordinate in the LLM coordinate system is also quantized, \ie
$\Delta y\sim l^2$. This latter result is not an outcome of  the
LLM geometry setup, or a direct consequence of the corresponding
dual $\N=4$ SYM analysis. (In the finite $J$, $y$ also has a
finite extent and it has been cut off at $R^2/2=l^2 J/2$.)
Moreover, if we follow the same line of arguments we find that the
function $z$ in the LLM geometries \cite{LLM} should also be
quantized, and at arbitrary $y$, $z$ should only take fractional
values, ranging from $-1/2$ to $+1/2$ \cite{Mosaffa-prog}.
\section{Less BPS Solutions}
\label{Less-BPS-section}%
Having briefly studied 1/2 BPS configurations, now we investigate
less BPS configurations including those of 1/4 and 1/8 BPS
solutions. Following the strategy explained in section
\ref{section2.3}, we should find the appropriate projector $\P$
for solving \eqref{variation1} and \eqref{variation2} as well as
the Gauss law constraint. In this section we will focus on the
projectors which in general keep 1/16 or 1/8 of supersymmetry.
Since our fermions carry two sets of fermionic indices we should
in principle introduce two projectors for each of the two  $SO(4)$
subspaces. In section \ref{section4.1} we study configurations
coming from deformations of the 1/2 BPS fuzzy sphere solutions of
the previous section. Here we have 1/8 and 1/4 BPS states. Within
our construction the 1/4 BPS states are obtained as special cases
of 1/8 BPS states and 1/2 BPS configurations as special case of
1/4 BPS. In section \ref{section4.2} we study 1/8 BPS
configurations which are not related to 1/2 BPS fuzzy sphere
solutions for any value of the parameters defining the solutions.
In section \ref{section4.3} we double check our BPS analysis by
working out the energy and other quantum numbers of the
configurations and directly verify the BPS condition using the
right hand side of the superalgebra.

\subsection{Multi spin 1/2 BPS spherical branes}\label{section4.1}
Here we identify a class of configurations which keep 2 or 4
number of supercharges. All the solutions of this class physically
correspond to rotating 1/2 BPS spherical three branes of the
previous section. In section \ref{section4.1.1} we work out and
analyze 1/8 BPS configurations and in section \ref{section4.1.2}
we study 1/4 BPS configurations.

\subsubsection{1/8 BPS configurations}\label{section4.1.1}
To obtain 1/8 BPS configurations coming from deformations of fuzzy
three spheres, let us start with the case in which we have a
single fuzzy three sphere in the $X^i$ directions, setting
$X^{a}=P^{a}=0$. The appropriate projection is%
\be\label{1/8BPS-projector1}%
\bar\P = \frac{1}{2}(1+ i\bar\sigma^{12}) \qquad {\rm OR}\qquad \P
= \frac{1}{2}(1+i\sigma^{12}). \ee%
Note that these projectors only act on the first index of our
fermions, \ie the Weyl indices of $SO(4)_i$. $\Tr\bar\P$ or
$\Tr\P$ over the space of Weyl indices of $SO(4)_i\times SO(4)_a$
equals to $4$ and hence we can describe 1/8 BPS states with either
of these projectors. Plugging $\bar\P$ into the fermion SUSY
variations we obtain%
\be\label{BPS-barP} %
\delta_\epsilon(\theta_{\alpha\beta})_{rs} =
\Big(P^l+i\frac{\mu}{R_-}X^l +
\frac{i}{3!g_s}\epsilon^{ijkl}[X^i,X^j,X^k,{\cal L}_5]\Big)_{rs}
(\sigma^l)^{\dot\gamma}_\alpha\
\frac{1}{2}(\delta_{\dot\gamma}^{\dot\eta} +
i\sigma^{12\dot\eta}_{\dot\gamma})\epsilon^0_{\dot\eta\beta}
\ee%
while $\delta_\epsilon(\theta_{\dot\alpha\dot\beta})$ can never
become zero. If we choose $\P$,
$\delta_\epsilon(\theta_{\alpha\beta})$ is non-vanishing and%
\be\label{BPS-P}%
\delta_\epsilon(\theta_{\dot\alpha\dot\beta})_{rs} =
\Big(P^l+i\frac{\mu}{R_-}X^l +
\frac{i}{3!g_s}\epsilon^{ijkl}[X^i,X^j,X^k,{\cal L}_5]\Big)_{rs}
(\sigma^l)^\gamma_{\dot\alpha}\ \frac{1}{2}(\delta_{\gamma}^{\eta}
+ i\sigma^{12\eta}_{\gamma})\epsilon^0_{\eta\dot\beta}%
\ee%
Expanding and setting coefficient of different $\sigma$'s and
$\bar{\sigma}$'s to zero independently, we have%
\bse\label{P-1/8BPS}
\begin{align} \label{con1}(P^1+i\frac{\mu}{R_-}X^1) -
i(P^2+i\frac{\mu}{R_-}X^2) - \frac{i}{g_s}[X^2,X^3,X^4,{\cal
L}_5]+ \frac{1}{g_s}[X^1,X^3,X^4,{\cal L}_5] &= 0, \\ \label{con2}
(P^3+i\frac{\mu}{R_-}X^3) - i(P^4+i\frac{\mu}{R_-}X^4) -
\frac{i}{g_s}[X^1,X^2,X^4,{\cal L}_5]+
\frac{1}{g_s}[X^1,X^2,X^3,{\cal L}_5] &= 0,
\end{align}%
\ese%
for the projector $\bar\P$ (resulting form \eqref{BPS-barP}) and%
\bse\label{barP-1/8BPS}\begin{align}%
\label{con3}(P^1+i\frac{\mu}{R_-}X^1) - i(P^2+i\frac{\mu}{R_-}X^2)
- \frac{i}{g_s}[X^2,X^3,X^4,{\cal L}_5]+
\frac{1}{g_s}[X^1,X^3,X^4,{\cal L}_5]
&= 0,\\ %
\label{con4}(P^3+i\frac{\mu}{R_-}X^3) + i(P^4+i\frac{\mu}{R_-}X^4)
- \frac{i}{g_s}[X^1,X^2,X^4,{\cal L}_5] -
\frac{1}{g_s}[X^1,X^2,X^3,{\cal L}_5] &=
0, \end{align}%
\ese%
resulting from \eqref{BPS-P} for the other possible projector
$\P$.

Looking for 1/8 BPS configurations we should only consider one set
of the above equations. For 1/4 BPS states, however, (4.4) and
(4.5) should be solved simultaneously, this will be done in
further detail in section \ref{section4.1.2}. From now on we only
focus
 on \eqref{con1}, \eqref{con2}. These complicated looking
equations are simplified drastically once we fix the
\eqref{gauge-fixing} gauge:%
\bse\begin{align}
(\dot{X^2}+i\dot{X^1})+\frac{i\mu}{R_-}(X^2+iX^1)=0, \\
(\dot{X^4}+i\dot{X^3})+\frac{i\mu}{R_-}(X^4+iX^3)=0,
\end{align}\ese%
whose solution are%
\bse\label{rotating-frame}
\begin{align}
(X^2+iX^1)=(X^2_0+iX^1_0)\exp(-\frac{i\mu}{R_{-}}t), \\
(X^4+iX^3)=(X^4_0+iX^3_0)\exp(-\frac{i\mu}{R_{-}}t).\end{align}%
\ese%

In this gauge,  the Gauss law takes a non-trivial form%
\be\begin{split}\label{gauss-law-1/8BPS}%
& \frac{2\mu}{R_-}[X^1,X^2] +
\frac{1}{g_s}\Bigl([X^1,[X^1,X^3,X^4,{\cal L}_5]] +
[X^2,[X^2,X^3,X^4,{\cal L}_5]]\Bigr) +\cr &
\frac{2\mu}{R_-}[X^3,X^4] +
\frac{1}{g_s}\Bigl([X^3,[X^1,X^2,X^3,{\cal L}_5]] +
[X^4,[X^1,X^2,X^4,{\cal L}_5]]\Bigr) = 0 .%
\end{split}\ee%
It is readily seen that the time dependence in the Gauss law drops
out and  \eqref{gauss-law-1/8BPS} reduces to an equation among
$X_0^i$'s. To simplify the notation we will use $X^i$ instead of
$X^i_0$'s. The main task is now to solve \eqref{gauss-law-1/8BPS}.

To solve \eqref{gauss-law-1/8BPS} we note that all the geometric
fluctuations of a three sphere can be expanded in terms of $SO(4)$
spherical harmonics, and hence one may try the ansatz
\be%
X^i \sim \ T^i\ _{i_1i_2\cdots i_n}\
\J^{i_1}\J^{i_2}\cdots\J^{i_n}%
\ee%
where $T$ is a tensor of $SO(4)_i$ of rank $n+1$.\\
{\it Conventions:}%

Hereafter, unless explicitly specified, we take $\J$'s to be in
$J\by J$ irreducible representation of $SO(4)$. In our notation
$\sim$  means that $X^i$'s are measured in units of
$l=\big(\frac{\mu g_s}{R_-}\big)^{1/2} l_s$. That is, $X^i\sim
\J^i$ means $X^i=l\J^i$.

In general one may search for solutions with arbitrary rank $n$.
In the present work we restrict ourselves to $n=1$. The general
case will be considered elsewhere \cite{TGMPT}. For the $n=1$
case,
we have \be\label{linear-J}%
X^i \sim \ T^i\ _j\ \J^j.%
\ee%
Plugging \eqref{linear-J} into  \eqref{gauss-law-1/8BPS} and using
\eqref{J-def}, the Gauss law equation reduces to simple algebraic
equation for the tensor $T$. Recalling that we also have the
${\cal L}_5$ in our disposal, \eqref{linear-J} is not the most
general
$X^i$'s at $n=1$ level, it is rather%
\be\label{general-linear-J}
 X^i \sim\ M^i\,_j\ \J^j + N^i\,_j\ i\J_5\J^j%
\ee%
where $M$ and $N$ are $4\by 4$ $SO(4)$ covariant matrices. Of
course not all  $M$ and $N$ matrices lead to physically distinct
configurations. For example, $({\tilde M},\ \tilde{N})$%
\be\label{MN-Ualpha}\begin{split}%
\tilde{M}^i\,_j&=\cos2\alpha M^i\,_j+\sin2\alpha N^i\,_j\\
\tilde{N}^i\,_j&=\cos2\alpha N^i\,_j-\sin2\alpha M^i\,_j
\end{split}\ee%
and $(M,N)$, which are related by a $U_{\alpha}$ transformation,
are physically equivalent. (Recall that fixing the gauge by
\eqref{gauge-fixing} the $U_\alpha$ rotations remains unfixed.) To
obtain \eqref{MN-Ualpha} we have used the fact that $\{\J^i,\J_5\}
= 0$ ({\it cf.} \eqref{J-L5-anticommute}).

To classify all possible $(M,N)$ configurations it is useful to
think them as exponentials of elements of $so(4)_i$ algebra. The
$4\by 4$ matrices are then related to symmetric traceless,
anti-symmetric, or singlet irreducible representations of $so(4)$.
The symmetric $M$ which can be written as $K-\frac{1}{4}(\Tr
K){\bf 1}$, for some symmetric matrix $K$, corresponds to
symmetric traceless representation. $M\propto {\bf 1}$ is of
course the singlet representation and if $M$ is in the form of
$SO(4)$ rotation matrix it can be written as the exponential of an
anti-symmetric $so(4)$ tensor. Let us, for the time being, set
$N=0$. It is evident that the anti-symmetric representation for
$M$ does not correspond to a physical deformation of the fuzzy
three sphere, it corresponds to zero mode fluctuations
\cite{TGMT}. The singlet representation, which is related to the
breathing mode of a three sphere brane does not solve the BPS
equation (similar statements is also true for membranes of the BMN
matrix model \cite{DSV1, DSV2}). Therefore, for the
$N=0$ the only remaining option is choosing $M$ to be of the form%
\[
M^i\,_j = \left(\begin{array}{cccc} a & 0 & 0 & 0 \\ 0 & b & 0 & 0 \\
0 & 0 & c & 0 \\ 0 & 0 & 0 & d \end{array}\right)\]%
where we have used six independent $SO(4)$ rotations to bring the
$M$ into the diagonal form.

It is straightforward to check that the Gauss law
\eqref{gauss-law-1/8BPS} is solved with
\be \label{ansatz}%
X_1\sim a\J_1,\quad X_2\sim b\J_2,\quad X_3\sim
c\J_3,\quad X_4\sim d\J_4%
\ee%
provided that%
\[
  (a^2+b^2)\frac{cd}{ab} = 2\ ,\qquad
 (c^2+d^2)\frac{ab}{cd} =2,
 \]%
or equivalently%
\be\label{abcd}
\frac{1}{a^2}+\frac{1}{b^2}  = \frac{1}{c^2}+\frac{1}{d^2}\
,\qquad \frac{a}{b}+\frac{b}{a}=\frac{2}{cd}\ .
 \ee%
The solutions to \eqref{abcd} can be parameterized by the angles
$\theta,\ \phi$:%
\be\label{theta-phi}%
a^{-1} = \kappa\sin\theta,\quad b^{-1} = \kappa\cos\theta,\quad
c^{-1} = \kappa\sin\phi,\quad d^{-1} =
\kappa\cos\phi%
\ee%
where%
\be\label{kappa}%
\kappa=\Big(\frac{1}{2}\sin2\theta\sin2\phi\Big)^{-1/2}\ .%
\ee%
For $\theta=\phi=\pi/4$, $a=b=c=d=1$, and hence the above solution
reduces to the half BPS spherical solution. This is of course
expected, because this family of 1/8 BPS configurations should
contain 1/2 BPS ones as special cases. If we expand the solution
about  $\theta=\phi=\pi/4$ as $\theta=\pi/4+\delta,\
\phi=\pi/4+\epsilon$, $\det M=abcd=1+{\cal
O}(\epsilon^2,\delta^2)$. This confirms the expectation that the
above BPS modes are coming from the exponentials of symmetric
traceless $so(4)$ representations.

As we have already mentioned, the above solution is describing
geometric fluctuations (deformations) of a  spherical three brane.
The shape of the deformed three brane, in the frame which is
rotating in both $X^1+iX^2$ and $X^3+iX^4$ directions with
frequency $\mu/R_-$ ({\it cf.} \eqref{rotating-frame}),
 can be easily worked out:%
\[
\frac{1}{a^2}X^2_1+\frac{1}{b^2}X^2_2+\frac{1}{c^2}X^2_3+\frac{1}{d^2}X^2_4=R^2{\bf
1}=l^2J\
{\bf 1}_{J\by J}%
\]
or equivalently%
\be\label{shape-abcd}%
\sin^2\theta X^2_1+ \cos^2\theta X^2_2+\sin^2\phi X^2_3+\cos^2\phi
X^2_4= R^2/\kappa^2
\ee%
In the large matrices limit the above equation is describing a
three brane of an ellipsoidal form. Although generic cross
sections of the above ellipsoid is like an ellipse, it has two
independent circular sections, e.g. along 2-plane
\[
X_3=A X_1\ ,\qquad X_4=0
\]
where $A^2=\frac{\cos2\theta}{\sin^2\phi}$
and 2-plane%
\[X_2=0,\qquad X_1=B X_3 \]%
where $B^2=\frac{\cos2\phi}{\sin^2\theta}$. In both cases the
section is circular with radii squares $R_1^2=R^2
\tan\theta\sin2\phi$ and $R_2^2=R^2 \tan\phi\sin2\theta$,
respectively. This configuration, hence, generically preserves
$U(1)\times U(1)$ isometry out of the $SO(4)$. There are special
cases with larger isometry subgroup which will be discussed in
section \ref{section4.1.2}.

The total energy and the angular momentum for this configuration
can be evaluated once we plug the ansatz \eqref{ansatz} into
\eqref{Hamiltonian}, \eqref{Jij}%
\be%
{\bf H} =
\frac{\mu^{2}l^2}{4R_{-}}\bigg((a-bcd)^{2}+(b-acd)^{2}+(c-abd)^{2}+(d-abc)^{2}\bigg)\
\Tr(\J^{2})%
\ee%
\be%
{\bf J}_{12} = \frac{\mu l^2}{4R_{-}}(a^{2}+b^{2}-2abcd)\ \Tr(\J^{2})%
\ee%
\be%
{\bf J}_{34} =
\frac{\mu l^2}{4R_{-}}(c^{2}+d^{2}-2abcd)\ \Tr(\J^{2})%
\ee%
and other angular momentum components vanish. (Note that we have
used the fact that
$\Tr(\J_1^2)=\Tr(\J_2^2)=\Tr(\J_3^2)=\Tr(\J_4^2)=\frac{1}{4}\Tr(\J^2)=J^2/4$).
Using \eqref{abcd}, energy
of the 1/8 BPS configurations becomes%
\be\label{1/8BPS-energy}%
{\bf H} =\frac{\mu}{2 g_{eff}^2}\
\bigg(\frac{\sin2\phi}{\sin2\theta}
+\frac{\sin2\theta}{\sin2\phi}-2\sin2\theta\sin2\phi\bigg)%
\ee%
where as discussed in \cite{TGMT, hedge-hog, squashed}%
\be\label{g-eff}%
g^2_{eff}=\left(\frac{R_-}{\mu
J}\right)^2\frac{1}{g_s}=\frac{1}{(\mu p^+)^2
g_s}%
\ee%
 is the effective coupling of the field theory on a giant
graviton of radius $R$, $R^2/l_s^2=\mu p^+ g_s$ \eqref{radius-J}.
This result is expected if the above deformations are
parameterizing the field excitations of an effective (Yang-Mills)
gauge theory of coupling $g_{eff}$ which resides on a giant
graviton. Similar results was obtained for deformations of the
giant in a background NSNS $B$-field, or when the magnetic field
on the brane is turned on \cite{squashed}.
The angular momenta is obtained to be%
\be\label{J12-J34}%
\begin{split}
{\bf J}_{12}&=\frac{1}{2g^2_{eff}}\
\bigg(\frac{\sin2\phi}{\sin2\theta}-\sin2\theta\sin2\phi\bigg)\\%
{\bf J}_{34}&=\frac{1}{2g^2_{eff}}\
\bigg(\frac{\sin2\theta}{\sin2\phi}-\sin2\theta\sin2\phi\bigg) .
\end{split}
\ee%
As we see and expected under the exchange of  $\theta$ and $\phi$
${\bf J}_{12}$ goes over to ${\bf J}_{34}$ while the expression
for energy is invariant.

Here we have only considered the classical matrix theory, if we
performed  computations with the quantized matrices we should
obtain quantized values for angular momenta. Imposing
semi-classical quantization of the angular momenta by hand we see
that the deformation parameters, $\theta, \phi$ cannot take
continuous values and become quantized as well.

Deformations of a three brane (giant graviton) besides the
geometric modes that we have considered also contains internal
``photon'' modes which one would expect to be in the same
super-multiplet as the geometric fluctuations \cite{hedge-hog}.
These photon modes are also among the 1/8 (and also possibly 1/4)
BPS states. As we showed the $N=0$ deformations only include the
geometric modes. To see the photon modes one should then consider
 $N\neq 0$ cases. To simplify the argument let us only
consider the infinitesimal deformations of the giant from the
spherical case. Since we only want to focus on the photonic modes,
or the ${\cal L}_5$ piece, we may take $M={\bf 1}$, \ie
$X^i=\J^i+iN^i\,_j{\cal L}_5\J^j$. One can show that among  the
three possibilities for $N$, only those which are in the
anti-symmetric representation of $so(4)$ lead to independent
(non-geometric) modes. The computations may be performed using the
identities presented in Appendix A. Since they are tedious and not
illuminating we will not present them here. However, here is a
simple intuitive argument for this statement. The geometric
fluctuations (which are linear in $\J_i$) will all show up in the
equation defining the shape, $g_{ij} X^iX^j\propto {\bf 1}_{J\by
J}$ where $g_{ij}$ is a symmetric $4\by 4$ matrix. It is now easy
to check that for $X^i\sim \J^i+iN^i\,_j {\cal L}_5\J^j$ where $N$
is the infinitesimal deformation parameter, the anti-symmetric
part of $N$, $F_{ij}$, is not completely captured in the shape
equation. For anti-symmetric $F$, $g_{ij}X^iX^j$ has a term of the
form $iF_{ij} {\cal L}_5 [\J^i,\J^j]$. This $F_{ij}$ may directly
be related to the gauge field strength living on the
brane.\footnote{Recall that to obtain the TGMT action we
discretized the brane action in which the gauge field was turned
off. In the discretization procedure, however, we introduced the
${\cal L}_5$. It is now apparent that introduction of the ${\cal
L}_5$ is crucial for reviving the gauge field which lives on the
three brane giants. Needless to emphasize that presence of this
photon modes and gauge fields is necessary to have a
supersymmetric theory.} Using $SO(4)$ rotations any anti-symmetric
$so(4)$ tensor $F_{ij}$  can always be brought to the skew
symmetric form%
\be\label{Fij}%
F_{ij} = \left(\begin{array}{cccc} 0 & f_1 & 0 & 0 \\ -f_1 & 0 & 0 & 0 \\
0 & 0 & 0 & f_2 \\ 0 & 0 & -f_2 & 0 \end{array}\right)%
\ee%
where $f_1$ and $f_2$ are the two parameters corresponding to the
two polarizations of photons.

In sum, the infinitesimal physical BPS deformations of three
sphere giants which are linear in $\J^i$'s can be parameterized as
$X^i=\J^i+S^i\,_j \J^j+iF^i\,_j{\cal L}_5\J^j$ where $S$ is a
symmetric traceless and $F$ an anti-symmetric $so(4)$ tensor. This
result is compatible with those of continuum limit
\cite{hedge-hog}.

In this section we considered the single giant configurations.
Multi giant configurations, where $\sum_{i}^4 (\J^i)^2$ is block
diagonal can also be treated in a very similar manner and hence we
do not repeat the details of computations here. For example for a
double giant configuration, one has two possibilities: to put the
deformations on either of the giants and keep the other one
intact, or generically deform both of the giants in such a way
that each of them individually satisfy the BPS equation
\eqref{abcd}. In the language of the notations we used here, that
is deforming each giant with a set of $\theta, \phi$ parameters.

\subsubsection{1/4 BPS configurations}\label{section4.1.2}
As mentioned earlier the BPS configurations resulting  from the
projectors $\P$ {\it or} $\bar\P$ \eqref{1/8BPS-projector1}, \ie
solutions to \eqref{P-1/8BPS} {\it or} \eqref{barP-1/8BPS},
preserve 4 supersymmetries, the first preserving two of
$Q_{\alpha\dot\beta}$ type supercharges while killing all
$Q_{\dot\alpha\beta}$ supercharges and the other keep two of
$Q_{\dot\alpha\beta}$, killing $Q_{\alpha\dot\beta}$ type
supercharges. It is, however, possible to impose both of the
projectors simultaneously and obtain a configuration which
preserves 8 supercharges.

To obtain such 1/4 BPS configurations one then needs to find
solutions to all four \eqref{con1}, \eqref{con2}, \eqref{con3} and
\eqref{con4} equations. It is readily seen that such solution must
be static in $34$ directions, that is ${\bf J}_{34}=0$. Moreover,
as solutions to \eqref{con1}, \eqref{con2}, one may use
\eqref{J12-J34} and impose the ${\bf J}_{34}=0$ condition. This is
possible if $\phi=\frac{\pi}{4}$ or $\frac{3\pi}{4}$. Therefore,
quarter BPS solutions constitute a one parameter family of the
rotating spherical branes with%
\be\label{1/4BPS-energy-J12}%
{\bf H} \equiv \mu {\bf J}_{12} =
\frac{\mu}{2g^2_{eff}}\big(\frac{1}{\sin2\theta}
-\sin2\theta\big),%
\ee%
which is positive definite and is zero for $\theta =\pi/4$ where
we recover 1/2 BPS configurations. One can also work out the shape
of the 1/4 BPS rotating branes:%
 \be\label{shape-1/4}%
 2\sin^2\theta X_1^2 + 2\cos^2\theta X_2^2 +
X_3^2 + X_4^2 = R^2 \sin2\theta%
\ee%
First we note that it has a two sphere cross section. To see this
set $X_1=r\sin\alpha$ and $X_2=r\cos\alpha$. For $\alpha=\pi/4$ we
recover a 2-sphere of radius squared $R^2\sin2\theta$ in
$r34$-space. This exhibits the $SU(2)$ isometry of the solution.
There is a circular cross section e.g. at $X_4=0$, $X_3=AX_1$
$(A^2=2\cos\theta)$, with radius squared $R^2\tan2\theta$.
Therefore, altogether for 1/4 BPS configuration we have
$SU(2)\times U(1)$ isometries out of whole $SO(4)$.

For the multi giant case, the configuration which is made of
several concentric 1/4 BPS states of the same kind (\ie e.g both
have vanishing ${\bf J}_{34}$)  still remains 1/4 BPS. However, if
we have concentric 1/4 BPS states of different kind, e.g. one of
them has a vanishing ${\bf J}_{34}$ and the other vanishing ${\bf
J}_{12}$, the system altogether is a 1/8 BPS configuration.

We discussed that there is another class of 1/8 BPS states which
are not of the form of  geometric fluctuations of a spherical
brane, rather related to turning on a gauge field on the brane.
These states are specified by an anti-symmetric rank two tensor of
$SO(4)$ $F_{ij}$, that is a ${\bf 6}$ of $SO(4)$. In terms of the
$SU(2)\times SU(2)$ it is $({\bf 3, 1})\oplus ({\bf 1, 3})$. If we
take the self-dual (or anti-self dual) part of $F_{ij}$, in the
notation of \eqref{Fij} that is $f_1=f_2$ (or $f_1=-f_2$), we will
have 1/4 BPS state. As is obvious, for these cases one of the
$SU(2)$'s, in which the $F_{ij}$ is a singlet, remains intact and
hence the symmetry is $SU(2)\times U(1)$.

\subsection{Other 1/8 BPS configurations}\label{section4.2}
So far we have analyzed configurations which could be obtained as
(continuous) deformations of spherical 1/2 BPS three brane giants.
In this section, we consider cases which are not of this kind. We
discuss two examples. One of them, the hyperboloid three brane,
only exists in the infinite $J$ limit while the other are
generically of the form of deformed (rotating) spherical branes
which are extended in $12$ and $56$ directions.

\subsubsection{Hyperboloid case}\label{section4.2.1}
Another class of solutions can be extracted out of matrices
$\K_i$ which satisfy the following algebraic structure%
\bea\label{Ki-brackets}
 [\K_1,\K_2,\K_3,{\cal L}_5] = -\K_4 \ ,&\quad
[\K_1,\K_2,\K_4,{\cal L}_5] = -\K_3 \cr [\K_1,\K_3,\K_4,{\cal L}_5] = -\K_2\ ,&\quad [\K_2,\K_3,\K_4,{\cal L}_5] = -\K_1. \eea%
Here $\K_i$'s are in the representations of $SO(2,2)$, rather than
$SO(4)$ ($[\K_i,\K_j]$ form generators of $so(2,2)$). It is easy
to check
that the Gauss law can be solved by%
\be X_1\sim a\K_1,\quad X_2\sim b\K_2,\quad X_3\sim c\K_3,\quad
X_4\sim d\K_4\ee%
provided that%
\be (a^2-b^2)\frac{cd}{ab} =2\ ,\qquad (c^2-d^2)\frac{ab}{cd} =2
\ee%
or equivalently%
\be \frac{1}{b^2}-\frac{1}{a^2}= \frac{1}{d^2}-\frac{1}{c^2}\
,\qquad \frac{a}{b}-\frac{b}{a}=\frac{2}{cd}\ee The solution can
be identified by $\theta,\phi$ as%
\be\label{solution2} a^{-1} = \kappa\sinh\theta,\quad b^{-1} =
\kappa\cosh\theta,\quad c^{-1} = \kappa\sinh\phi,\quad d^{-1} =
\kappa\cosh\phi\ee%
where%
\be\kappa = \Big(\frac{1}{2}\sinh2\theta\sinh2\phi\Big)^{-1/2} \ee%
In this case, the Casimir operator is $\K^2 =
-\K_1^2+\K_2^2-\K_3^2+\K_4^2$ and therefore, the shape can be
described by%
\be%
-\sinh^2\theta X_1^2+\cosh^2\theta
X_2^2-\sinh^2\phi X_3^2+\cosh^2\phi X_4^2=R^2/\kappa^2%
\ee%
The above describes a hyperboloid which is extended off to
infinity in all directions. It has two independent circular cross
sections while generically we also have hyperbolic sections as
well. This configuration generically  preserves $U(1)\times U(1)$
symmetry out of the $SO(4)$. As we can see there is no real
$\theta$ and $\phi$ for which the above goes to a spherical brane.
 Similar
configurations (hyperbolic membrane solutions) in the BMN matrix
model, i.e. the tiny membrane graviton matrix theory, has been
constructed in \cite{Bak}.

 It is straightforward to see that the above solutions
\emph{formally} could be obtained from those of section
\ref{section4.1} by Wick rotation on $\theta$ and $\phi$. Doing
so, noting that $\J$'s are hermitian the corresponding $\K$'s
cannot be hermitian. The hermiticity problem of $X^i$ is resolved
once we note that  $so(2,2)$, unlike $so(4)$, is non-compact and
has no finite dimensional unitary representations.\footnote{Using
the Wick rotation trick it can be checked that hermiticity of the
four bracket equations \eqref{Ki-brackets} forces us to Wick
rotate $\theta$ and $\phi$ together. Hence, it is not possible to
obtain a solution based on the generators of $so(3,1)$.}
Therefore, solutions of this class only exist for infinite $J$.
This is compatible with the intuition that an infinite size brane
cannot be built out of finite number of building blocks, the tiny
gravitons.   One should however, note that the infinite $J$ limit
should be taken with a special care to keep physical light-cone
moment finite, that is, $J,\ R_-\to \infty,\ \mu p^+=\frac{\mu
J}{R_-}=fixed$. In the large matrices limit, one can understand
this hyperbolic solutions are infinite deformation limit of the
spherical branes of the previous section.

Total energy and the angular momentum for this configuration can be evaluated%
\be {\bf H} = \frac{\mu^{2}l^2}{4R_{-}}
\bigg(-(a+bcd)^{2}+(b-acd)^{2}-(c+abd)^{2}+(d-abc)^{2}\bigg)\
\Tr(\K^{2})\ee%
\be {\bf J}_{12} = \frac{\mu l^2}{4R_{-}}
\big(b^{2}-a^{2}-2abcd\big)\ \Tr(\K^{2}) \ee%
\be {\bf J}_{34} = \frac{\mu l^2}{4R_{-}} \big(d^{2}-c^{2}-2abcd)\
\Tr(\K^{2}) \ee%
and other angular momentum component vanish. Note that in the
above equations, we have used the fact that\
$\Tr(\K_2^2)=\Tr(\K_4^2)=-\Tr(\K_1^2)=-\Tr(\K_3^2)=\frac{1}{4}\Tr(\K^2)$.
Using \eqref{solution2}, energy and angular momenta read%
\be\label{hyperbolic-energy}%
{\bf H} = \frac{\mu}{2 g_{eff}^2}\
\bigg(\frac{\sinh2\phi}{\sinh2\theta}
+\frac{\sinh2\theta}{\sinh2\phi}+2\sinh2\theta\sinh2\phi\bigg)%
\ee%
\be\label{hyperbolic-J12-J34}%
\begin{split}
{\bf J}_{12}&=\frac{1}{2g^2_{eff}}\
\bigg(\frac{\sinh2\phi}{\sinh2\theta}+\sinh2\theta\sinh2\phi\bigg)\\%
{\bf J}_{34}&=\frac{1}{2g^2_{eff}}\
\bigg(\frac{\sinh2\theta}{\sinh2\phi}+\sinh2\theta\sinh2\phi\bigg)
\end{split}
\ee%
where, despite of having infinite size matrices, $g_{eff}$ is
still given by \eqref{g-eff}, that is $g^{-2}_{eff}=(\mu p^+)^2
g_s$. From the energy expression it is readily seen that the
Hamiltonian is positive definite and its value is finite.

It is worth noting that, unlike the case of previous section, here
we do not have special cases in which we recover 1/4 BPS solutions
and where the $U(1)\times U(1)$ isometry enhances to $SU(2)\times
U(1)$. That is because ${\bf J}_{34}$ or ${\bf J}_{12}$ never
vanish while the other one remains finite.

\subsubsection{Deformed spherical branes in $X^1,X^2,X^5, X^6$
subspace}\label{section4.2.2}%
Let us now consider the case which
involves both $X^i$ and $X^a$ directions in a non-trivial way.
Take $X^{3}=X^{4}=X^{7}=X^{8}=0$ while the other $X$'s are turned
on. For this case
\eqref{variation1} takes the form%
\be\begin{split} \delta_\epsilon(\theta_{\alpha\beta}) &=
\bigg((\sigma^1)_{\alpha}^{\dot{\rho}}(P^1+\frac{i\mu}{R_-}X^1)\delta^\rho_\beta+
(\sigma^2)_{\alpha}^{\dot{\rho}}(P^2+\frac{i\mu}{R_-}X^2)\delta^\rho_\beta
\cr
&\qquad+\frac{1}{2g_s}(\sigma^1)_{\alpha}^{\dot{\rho}}(i\sigma^{56})_{\beta}^{\rho}[X^1,X^5,X^6,{\cal
L}_5] +
\frac{1}{2g_s}(\sigma^2)_{\alpha}^{\dot{\rho}}(i\sigma^{56})_{\beta}^{\rho}[X^2,X^5,X^6,{\cal
L}_5]\bigg)\epsilon^0_{\dot{\rho}\rho}\cr
&+\bigg((\sigma^5)_{\beta}^{\dot{\rho}}(P^5+\frac{i\mu}{R_-}X^5)\delta^\rho_\alpha+
(\sigma^6)_{\beta}^{\dot{\rho}}(P^6+\frac{i\mu}{R_-}X^6)\delta^\rho_\alpha\cr
&\qquad+\frac{1}{2g_s}(\sigma^5)_{\beta}^{\dot{\rho}}(i\sigma^{12})_{\alpha}^{\rho}[X^5,X^1,X^2,{\cal
L}_5] +
\frac{1}{2g_s}(\sigma^6)_{\beta}^{\dot{\rho}}(i\sigma^{12})_{\alpha}^{\rho}[X^6,X^1,X^2,{\cal
L}_5]\bigg)\epsilon^0_{\rho\dot{\rho}}
\end{split}\ee%
Similarly one may work out
$\delta_\epsilon(\theta_{\dot\alpha\dot\beta})$ using
\eqref{variation2}.

We choose our projector to be%
\be%
\P=\P_1\P_2,\qquad \P_1=\frac{1}{2}(1+i\sigma^{12})\ \ \ \ \P_2 =
\frac{1}{2}(1+
i\sigma^{56}) \ee%
Inserting projection and Expanding and setting coefficient of
different $\sigma$'s equal to zero independently, the BPS
equations are obtained to be
\bse\label{Rijab-BPS}\begin{align}
 (P^1+i\frac{\mu}{R_-}X^1) -
i(P^2+i\frac{\mu}{R_-}X^2) - \frac{i}{g_s}[X^2,X^5,X^6,{\cal
L}_5]+ \frac{1}{g_s}[X^1,X^5,X^6,{\cal L}_5] &= 0
\\ (P^5+i\frac{\mu}{R_-}X^5) -
i(P^6+i\frac{\mu}{R_-}X^6) - \frac{i}{g_s}[X^6,X^1,X^2,{\cal
L}_5]+ \frac{1}{g_s}[X^5,X^1,X^2,{\cal L}_5] &= 0
\end{align}\ese%
These are similar to \eqref{con1}, \eqref{con2} if one exchanges
$X^5\longrightarrow X^3$, $X^6\longrightarrow X^4$ and hence could
be solved with the same trick (note that this exchange should also
be done in the gauge fixing condition \eqref{gauge-fixing}). That
is, one may insert $X^1\sim a\J_1,\ X^2\sim b\J_2,\ X^5\sim
c\J_3,\ X^6\sim d\J_4$ into \eqref{Rijab-BPS} where $a,b,c$ and
$d$ are of the form \eqref{theta-phi}.

There is, however, a subtlety here. It is easy to verify that $\Tr
\P=2$, rather than 4, which is needed for obtaining 1/8 BPS
configuration and it may seem that the above equations are
describing 1/16 BPS solutions. This issue is resolved once we note
that there is another projector, namely%
\be%
\bar\P=\bar\P_1\bar\P_2\ ,\qquad
\bar\P_1=\frac{1}{2}(1+i\bar\sigma^{12})\ \ \ \ \bar\P_2 =
\frac{1}{2}(1+ i\bar\sigma^{56})%
\ee%
which also leads to the same BPS equations as \eqref{Rijab-BPS}.
Therefore, solutions of \eqref{Rijab-BPS} will preserve both of
the supersymmetries resulting from $\P$ and $\bar\P$ projectors
and hence they are describing 1/8 BPS states.

Although very similar in equations, the above deformed three
sphere configurations are physically totally distinct  from those
of section \ref{section4.1.1}. For this class of solutions the
(central) extension $\R_{ijab}$ \eqref{R} does not vanish.
Parameterizing the solutions of \eqref{Rijab-BPS} by $\theta$ and
$\phi$, as is done in \eqref{theta-phi}, one can now calculate the
quantum numbers
associated with the above solutions:%
\bse\label{Rijab-QN}\begin{align}%
{\bf H} &=
\frac{\mu}{2g^2_{eff}}\bigg(\frac{\sin2\phi}{\sin2\theta}
+\frac{\sin2\theta}{\sin2\phi}\bigg)\\%
{\bf J}_{12}&=\frac{1}{2g^2_{eff}}\
\bigg(\frac{\sin2\phi}{\sin2\theta}-\sin2\theta\sin2\phi\bigg)\\%
{\bf J}_{56}&=\frac{1}{2g^2_{eff}}\
\bigg(\frac{\sin2\theta}{\sin2\phi}-\sin2\theta\sin2\phi\bigg) \\
\R_{1256}&=\frac{1}{g_s}\Tr([X^1,X^2,X^5,X^6]{\cal L}_5) =
+\frac{1}{4g_{eff}^2}\sin2\theta\sin2\phi \end{align}\ese%

It is note worthy that among these solutions we have 1/4 BPS
configurations in special case of ${\bf J}_{12}={\bf J}_{56}=0$
which happens for $\theta=\phi=\pi/4$, $\R_{1256}$ has its maximal
value. For this case we have spherical three branes. Out of the
$SO(4)_i\times SO(4)_a$ symmetry this configuration preserves
$SO(4)_{diag}$ and two extra $U(1)$'s. The radius of this sphere
if $\J$'s are in the irreducible representations of $SO(4)$ is
equal to $R$ \eqref{radius-J}. As discussed in \cite{extensions}
non-zero $\R_{1256}$ corresponds to the (self-dual) four form
dipole moment of the spherical three brane.

\subsection{Analysis from the superalgebra}\label{section4.3}
In the previous sections we focused on the definition of the BPS
states (configurations) which is resulting form the vanishing of
supersymmetry variations of the fermions. One can, however,
equivalently use the right-hand-side (RHS) of the superalgebra to
identify the BPS states. That is, the configurations which have
specific relations between their energy and other quantum charges
for which the RHS of the supercharge anticommutators vanish are
BPS. The number of supersymmetries preserved is then equal to the
number of the zero eigenvalues the RHS of the superalgebra has. We
analyze the cases of those only in $X^i$ subspace and in $X^i,
X^a$ subspaces separately in sections \ref{section4.3.1} and
\ref{section4.3.2}.

\subsubsection{Deformed 1/2 BPS cases}\label{section4.3.1}
This case consists of the configurations which have vanishing
$X^a, P^a$, and hence for these configurations ${\bf J}_{ab}=0$
and $\R_{ijab}=0$.  The RHS of the both of the $PSU(2|2)$
superalgebra factors are%
 \bse\label{HJ12J34}\begin{align}
\{Q_{\alpha \dot\beta},Q^{\dagger\rho \dot\gamma}\} &=
\delta^{\dot\beta}_{\dot\gamma}\left(\delta_{\alpha}^{\rho}\ {\bf
H} + \mu(i\sigma^{12})_{\alpha}^{\rho}\
{\bf J}_{12} + \mu(i\sigma^{34})_{\alpha}^{\rho}\ {\bf J}_{34} \right) \\
\{Q_{\dot\alpha\beta},Q^{\dagger\dot\rho\gamma}\}
&=\delta_{\beta}^\gamma\left( \delta_{\dot\alpha}^{\dot\rho}\ {\bf
H} + \mu(i\bar\sigma^{12})_{\dot\alpha}^{\dot\rho}\ {\bf J}_{12} +
\mu(i\bar\sigma^{34})_{\dot\alpha}^{\dot\rho}\ {\bf J}_{34}\right)
\end{align}\ese%
It is convenient to choose the basis such that%
\be  i\sigma^{12} = i\sigma^{34} = \left(\begin{array}{cc} 1 & 0 \\
0 & -1 \end{array}\right) \qquad,\qquad i\bar\sigma^{12} =
-i\bar\sigma^{34} = \left(\begin{array}{cc} 1 & 0 \\ 0 & -1
\end{array}\right) \ee%
The demand of having BPS configurations leads to%
\be\label{H-JJ1} \left(\begin{array}{cc}
 {\bf H} & 0 \\  0 & {\bf H} \\
\end{array} \right) + \mu\left(\begin{array}{cc} + {\bf J}_{12} +
{\bf J}_{34} & 0 \\ 0 & -{\bf J}_{12}-{\bf J}_{34}
\end{array} \right)\ = 0 \ee%
and%
\be\label{H-JJ2} \left(\begin{array}{cc} {\bf H} & 0 \\  0 & {\bf H} \\
\end{array} \right) + \mu\left(\begin{array}{cc} + {\bf J}_{12} -
{\bf J}_{34} & 0 \\ 0 & -{\bf J}_{12}+{\bf J}_{34}
\end{array} \right)\ = 0 \ee%
Dealing with complex supercharges, one should remember that the
number of preserved real supersymmetries is twice the number of
the zero eigenvalues.

For the configuration with energy and angular momenta of
\eqref{1/8BPS-energy} and \eqref{J12-J34}, it is evident that
${\bf H}=\mu({\bf J}_{12}+{\bf J}_{34})$ and hence the
configuration is 1/8 BPS, as discussed in the previous section. As
we see for this case both of the preserved supercharges are coming
from a single $PSU(2|2)$.

The 1/4 BPS configurations in this class, as  is readily seen from
\eqref{H-JJ1} and \eqref{H-JJ2}, can only be obtained if either of
${\bf J}_{12}$ or ${\bf J}_{34}$ vanishes while the other one is
equal to the energy (up to a factor of $\mu$). This is exactly the
case for the configurations of section \ref{section4.1.2}. In this
case two of the supercharges are coming from one $PSU(2|2)$ and
two of them from the other $PSU(2|2)$. In a similar manner, half
BPS configurations must have  ${\bf J}_{12}={\bf J}_{34}=0$.

The hyperbolic case which discussed in \ref{section4.2.1} can also
be analyzed in the same way. For this case only non-vanishing
generators are ${\bf H}, {\bf J}_{12}$ and ${\bf J}_{34}$. For
this configuration as seen from \eqref{hyperbolic-energy},
\eqref{hyperbolic-J12-J34} satisfies ${\bf H}=\mu({\bf
J}_{12}+{\bf J}_{34})$ and hence is 1/8 BPS.

\subsubsection{Configurations with non-zero
$\R_{ijab}$}\label{section4.3.2}%
For the configurations lying in 1256 direction the algebra reads%
\bse\label{}\begin{align}
\label{}\{Q_{\alpha\dot\beta},Q^{\dagger\rho\dot\lambda}\} &=
\delta_{\alpha}^{\rho} \delta_{\dot\beta}^{\dot\lambda}{\bf H} +
\mu(i\sigma^{12})_{\alpha}^{\rho}
\delta_{\dot\beta}^{\dot\lambda}{\bf J}_{12} +
\mu\delta_{\alpha}^{\rho}
(i\bar\sigma^{56})_{\dot\beta}^{\dot\lambda}{\bf J}_{56} -
4\mu(i\sigma^{12})_{\alpha}^{\rho}
(i\bar\sigma^{56})_{\dot\beta}^{\dot\lambda}\R_{1256} \\
\label{}\{Q_{\dot{\alpha}\beta},Q^{\dagger\dot{\rho}\lambda}\} &=
\delta_{\dot{\alpha}}^{\dot{\rho}} \delta_{\beta}^{\lambda}{\bf H}
+ \mu(i\bar\sigma^{12})_{\dot{\alpha}}^{\dot{\rho}}
\delta_{\beta}^{\lambda}{\bf J}_{12} +
\mu\delta_{\dot{\alpha}}^{\dot{\rho}}
(i\sigma^{56})_{\beta}^{\lambda}{\bf J}_{56} -
4\mu(i\bar\sigma^{12})_{\dot{\alpha}}^{\dot{\rho}}
(i\sigma^{56})_{\beta}^{\lambda}\R_{1256} \end{align}\ese%
In a convenient basis for $\sigma$'s both of the above equations
take the same form as%
\be\label{Rijab-RHS} {\bf H}+ \mu(s_1{\bf J}_{12} + s_2{\bf
J}_{56} - 4s_1s_2 \R_{1256})=0 \ee where $s_1,s_2$ are taking $\pm
1$ values.
Therefore the configuration of section \ref{section4.2.2} which
has ${\bf H}=\mu \left({\bf J}_{12}+{\bf
J}_{56}+4\R_{1256}\right)$, satisfies \eqref{Rijab-RHS} for
$s_1=s_2=-1$ and is a 1/8 BPS configuration. It is notable that
each couple of the four preserved supercharges are coming from one
the $PSU(2|2)$ superalgebras. For the special case of ${\bf
J}_{12}={\bf J}_{56}=0$ the above equation finds a new set of
solutions for $s_1=s_2=+1$ and hence the preserved SUSY is doubled
leading to a 1/4 BPS solution.
\section{ BPS States in the Dual SYM Theory}
The TGMT is conjectured to be the DLCQ of type IIB string theory
on the $AdS_5\times S^5$ or the plane-wave backgrounds. As such,
one expects the BPS states of the TGMT analyzed in the previous
sections to have counterparts in the dual $\N=4$ $U(N)$ SYM
theory. In this section we construct {\it local} operators in the
$\N=4$ SYM which correspond to these BPS states.

The $\N=4$, $D=4$, $U(N)$ SYM theory is a superconformal field
theory and has a large supergroup, $PSU(2,2|4)$. All the gauge
invariant operators of the gauge theory fall into various
multiplets of unitary representations of this superconformal group
which may be labelled by the quantum numbers of the bosonic
subgroup: $SU(2,2)\times SU(4)\simeq SO(4,2)\times SO(6)$ with%
\be\label{Dynkin-labels}%
\begin{array}{cccc}SU(2,2)&\approx&SO(4,2)&\supset SO(4)\times
SO(2)\approx \overbrace{SU(2)\times
SU(2)}^{(s_+,s_-)}\times \overbrace{U(1)}^{\Delta}\\
\underbrace{SU(4)}_{[r_1,r_2,r_3]}&\approx&SO(6)&\supset
SO(4)\times SO(2)\approx \underbrace{SU(2)\times
SU(2)}_{(t,u)}\times \underbrace{U(1)}_{J}\\\end{array}%
\ee%

In order to explicitly write the operators we need to recall the
 field content  of the $U(N)$ SYM theory
which consists of a spin one gauge field $A_\mu$, four spin 1/2
Weyl fermion fields $\psi^I_\alpha$ and six spin zero scalar
fields $\phi^i$, all in the $N\by N$ representation of $U(N)$.
Under $SU(4)$ $R$-symmetry, $A_\mu$ is a singlet, $\psi$ is a {\bf
4}, and $\phi$ is a rank two anti-symmetric {\bf 6}. These fields
naturally fall into an (ultra) short representation of
$psu(2,2|4)$.

Gauge invariant operators are obtained by summing over all $U(N)$
indices of any combination of the covariant derivative $D_\mu =
\partial_\mu + i[A_\mu, .\ ]$, $\psi^I_\alpha$ and the three complex
scalars $X,Y, Z$ defined as%
\be X=\frac{1}{\sqrt2}(\phi^1+i\phi^2),\qquad
Y=\frac{1}{\sqrt2}(\phi^3+i\phi^4),\qquad
Z=\frac{1}{\sqrt2}(\phi^5+i\phi^6). \ee%
Among these operators there are the BPS operators whose scaling
dimension is protected and completely specified by their
$SO(4)\subset SO(4,2)$ representations and $R$-charges. Almost
always people use $\Tr$ over a product of $N\by N$ matrices to
form gauge invariant operators. For our purposes, where we are
dealing with the brane-type states, the giant gravitons, the
subdeterminant \cite{Vijay} or Schur polynomial bases are more
appropriate \cite{Antal}. Let us focus on the subdeterminant
basis, generalization of the discussions to the other basis is
straightforward.

\subsection{Systematics of BPS operators}%
Independently of the $\Tr$, subdeterminant or Schur polynomial
bases, one can study and analyze the BPS operators just by their
$SO(4,2)\times SO(6)$ quantum numbers, as these are the groups
appearing in the $psu(2,2|4)$ superalgebra. The classification of
these BPS operators has been extensively studied  in the literature
\cite{Dobrev,Ferrara,D'Hoker} but generically with some emphasis on
the trace basis. As most of the arguments are basis independent, we
will be very brief on that.

All physical operators of the $\N=4$ SYM fall into unitary (or
more generally  unitarizable) multiplets of $psu(2,2|4)$
superalgebra. There is, however, a one-to-one correspondence
between superconformal chiral-primary operators and unitary
superconformal multiplets; the superconformal chiral-primaries
appear as the highest weight state of the superconformal BPS
multiplet. Hence, the multiplets can be named after their highest
weight state. {\it We should, however, stress that here we only
analyze individual states/configurations and not the multiplets}.
The chiral-primary operators are only made out of scalar fields
$\phi^i$ and hence completely specified with their $SU(4)$
$R$-symmetry representation.

Systematic analysis show that there are four distinct classes
\cite{Ferrara,D'Hoker}:
\begin{itemize}
\item $\Delta = r_1+r_2+r_3$ \item $\Delta =
\frac{3}{2}r_1+r_2+\frac{1}{2}r_3 \geq
2+\frac{1}{2}r_1+r_2+\frac{3}{2}$ \item $\Delta =
\frac{1}{2}r_1+r_2+\frac{3}{2}r_3 \geq
2+\frac{3}{2}r_1+r_2+\frac{1}{2}$ \item $\Delta \geq$ Max[$
2+\frac{1}{2}r_1+r_2+\frac{3}{2}r_3 ;
2+\frac{3}{2}r_1+r_2+\frac{1}{2}$]
\end{itemize}
where $r_i$ are the Dynkin labels of $SU(4)$ irreducible
representation ({\it cf.} \eqref{Dynkin-labels}).
 The first 3 cases correspond to discrete series of
representation for which at least one of the supercharges commutes
with the primary operator. These states are BPS states. A given
BPS state of specific charges in general can be either in the
highest weight representation of a BPS multiplet, or a descendent
of another BPS state. The fourth case corresponds to continuous
series of representations, for which no supercharges commute with
the primary operator. They are referred to as non-BPS operators.

\subsubsection*{1/2 BPS operators}
Half-BPS operators sit in $[0,k,0], k\geq 2$ (Dynkin label)
representation of $R$-symmetry \cite{Ferrara}. These chiral
primaries are annihilated by half of the super-Poincare charges,
$Q$'s, and appear as the highest weight state of a short (1/2 BPS)
multiplet with spin ranging from 0 to 2. The chiral-primary states
of $R$-charge $k$ are totally symmetric traceless rank $k$ tensors
of $SU(4)$. In terms of the $\N=4$ fields,
the simplest such \emph{local} operator is ${\cal O}_k$:%
\be\label{chiral-primary}%
{\cal O}_k =\frac{1}{\sqrt{k!(N-k)!}}{\cal E}_{i_1i_2\cdots
i_k}^{j_1j_2\cdots j_k}
 : Z_{i_1}^{j_1} Z_{i_2}^{j_2}\cdots
Z_{i_k}^{j_k}:\
 \ee%
 where
\be%
{\cal E}_{i_1i_2\cdots i_k}^{j_1j_2\cdots j_k}\equiv
\epsilon_{i_1i_2\cdots i_ki_{k+1}\cdots i_N}\epsilon^{j_1j_2\cdots
j_k i_{k+1}\cdots i_N}.%
\ee%
\eqref{chiral-primary}  corresponds to a giant graviton of radius
$R^2\sim k$. Note that the indices on $Z^i_j$ are running from 1
to $N$ and are $U(N)$ indices.

The most general half-BPS operators of $R$-charge $J$ is a
multi-subdeterminant operator of the form%
\be%
:{\cal O}_{k_1}(x){\cal O}_{k_2}(x)\cdots{\cal O}_{k_n}(x):\
,\quad\qquad J = \sum_{i=1}^n k_i\ .%
\ee%
The above is describing a multi-giant configuration consisting of
$n$ concentric giants, whose radii squared sum to $J$. As is
evident all of these operators have $\Delta-J=0$.

In order to compare to the TGMT it is more convenient to label
them by $SO(4)\times U(1)_\Delta \times SU(2)\times SU(2)\times
U(1)_J$ subgroup of $SO(4,2)\times SO(6)$ and identify the
light-cone Hamiltonian as \cite{TGMT, Review}%
\[{\bf H}=\Delta-J .\]%
The chiral primary operators which are singlets of $SO(4)\times
SO(4)_R$ correspond to the zero energy half BPS solutions of the
TGMT \cite{half-BPS}, see also section \ref{half-BPS-section}. In
fact there is an exact one-to-one correspondence between the TGMT
1/2 BPS configurations and the $\N=4$ \emph{local} chiral-primary
operators.

\subsection{Less BPS operators}%
As discussed 1/2 BPS operators are those which are only made out
of one kind of the complex scalars, e.g. $Z$. The 1/4 (and 1/8)
BPS chiral operators are those which besides $Z$ also involve $Y$
(and $X,\ Y$). Here is a more systematic analysis.

\subsubsection{1/4 BPS operators}%
Quarter-BPS operators are the next simplest, killed by four
super-Poincare charges. 1/4 BPS chiral operators sit in Dynkin
label $[l,J-l,l], J\geq l\geq 2$ representation of the
$R$-symmetry. These states can appear either as descendent of a
chiral primary or as the highest weight of a 1/4 BPS multiplet in
which spin of the states ranges from 0 to 3. The latter is only
possible if we have multi (at least two) trace/subdeterminant
operator \cite{D'Hoker}.

For example  operators of the form%
\be%
{\hat {\cal O}}_{J, l} = 
{\cal E}_{i_1i_2\cdots i_{J+l}}^{j_1j_2\cdots j_{J+l}}\ :
X_{i_1}^{j_1} X_{i_2}^{j_2}\cdots
X_{i_l}^{j_l}Z_{i_{l+1}}^{j_{l+1}}\cdots Z_{i_{J+l}}^{j_{J+l}}:\
 \ee%
correspond to a deformed single giant, with ${\bf J}_{34}=0$ and
${\bf J}_{12}\propto l$. These operators have ${\bf
H}=\Delta-J=l$. For $l=0$ these operators reduce to the chiral
primaries of \eqref{chiral-primary}.  Operators of this kind has
been considered in \cite{Beren3}. One can consider
multi-subdeterminant operators. These are of the form%
\[
:{\hat{\cal O}}_{J_1,l_1}(x){\hat{\cal
O}}_{J_2,l_2}(x)\cdots{\hat{\cal O}}_{J_n,l_n}(x):\ ,\qquad
\sum_{i=1}^n J_i=J,\ \sum_{i=1}^n l_i=l ,  \]%
which is of the form of $n$ concentric deformed spherical giants.
Obviously some of the $l_i$'s can be zero.

There is another class of 1/4 BPS states which is obtained by
insertion of the covariant derivative, instead of $X$ into the
sequence of $Z$'s in the 1/2 BPS operator \eqref{chiral-primary}.
In order to obtain a 1/4 BPS states, however, we should insert a
self-dual gauge field strength ({\it cf.} discussions in the end
of section \ref{section4.1.2}).

\subsubsection{1/8 BPS operators}%
The last family of chiral operators are 1/8 BPS operators sit in
$[l-m,J-l,l+m], l\geq m\geq 2$ representation of $R$-symmetry and
are annihilated by two supercharges. These operators are
constructed from $J$ number of $Z$'s, $l$ number of $X$'s and
 and $m$ of  $Y$'s, e.g.%
\be\label{1/8-BPS-opt}%
{\tilde{\cal O}}_{J, l,m} = {\cal E}_{i_1i_2\cdots
i_{J+l+m}}^{j_1j_2\cdots j_{J+l+m}}\ :X_{i_1}^{j_1}
X_{i_2}^{j_2}\cdots X_{i_l}^{j_l}Y_{i_{l+1}}^{j_{l+1}}\cdots
Y_{i_{l+m}}^{j_{l+m}} Z_{i_{l+m+1}}^{j_{l+m+1}}\cdots
Z_{i_{J+l+m}}^{j_{J+l+m}}:
 \ee%
or one may have insertions of $X$'s or $Y$'s (and not both)
together with insertion of a self-dual $F_{\mu\nu}$. In the TGMT,
although possible, we did not specify configurations corresponding
to the latter case. ${\tilde{\cal O}}_{J, l,m}$ operators
correspond to states with numbers ${\bf J}_{12}\propto l$ and
${\bf J}_{34}\propto m$ and ${\bf H}=\Delta-J=l+m$, in the matrix
theory side. These operators are invariant under $SO(4)\times
U(1)\times U(1)$. Similarly to the previous cases we can construct
multi-subdeterminants, corresponding to multi giant states.

In section \ref{section4.2.1} we analyzed hyperbolic
configurations.
 The gauge theory operators corresponding to this configuration
can be obtained from operators of the form \eqref{1/8-BPS-opt} in
the appropriate $J, N\to\infty$ limit, keeping $\mu p^+$ fixed,
that is the BMN limit \cite{BMN, Review}. One should also scale
$l$ and $m$ such that $l/N, m/N$ remain finite. Intuitively, one
would expect that for large angular momenta the deformation of the
rotating three brane from the spherical shape is so large that it
deforms to a hyperboloid shape brane.
\section{Discussion and Outlook}\label{discussion-section}
In this paper we continued analysis of the Tiny Graviton Matrix
Theory (TGMT) by classifying 1/4 and 1/8 BPS states of the Matrix
theory. These are generically of the form of deformed three sphere
giant gravitons or rotating spherical branes. We then compared
these configurations with the states in the dual $\N=4$ $U(N)$ SYM
theory and explicitly constructed the \emph{local} operator
corresponding to each configuration. In this way we provided
further evidence in support of the TGMT conjecture.

In the $\N=4$ SYM theory, however, one can have BPS
\emph{non-local} operators. For example, there are 1/2 BPS Wilson
lines \cite{Wilson-lines}. The supergravity solutions dual to such
operators have also been constructed \cite{Lunin}. It is
interesting to find the description of these non-local operators
in the TGMT language. Presumably such a configuration, which
should have a manifest $SO(5)\times SO(3)$ isometry \cite{Lunin},
could be obtained from turning on the $X^i$'s and one of the
$X^a$'s, say $X^5$, as $X^i=\J^i$ and $X^5={\cal L}_5$. In this
way  we have the desired $SO(5)\times SO(3)$ symmetry. The
detailed analysis of this 1/2 BPS configuration is postponed to an
upcoming work \cite{Ali-Akbari}.

We did not present an explicit gauge theory description for the
configurations of section \ref{section4.2.2} which involves
$\R_{ijab}$, and cannot be obtained along the ideas discussed here
in section 5, \ie by insertion of other fields into sequence of
chiral primaries. As they involve turning on a central extension
in the extended $psu(2,2|4)$ superalgebra \cite{extensions} one
would expect that these operators should also correspond to
non-local gauge theory operators. Identifying these operators, and
operators corresponding to the other possible (central) extensions
of the $psu(2,2|4)$ is another interesting direction to pursue.

Here we only studied transverse three branes, mainly those which
are topologically spherical ones. As another interesting direction
for further analysis, one can study BPS configurations
corresponding to longitudinal three branes or longitudinal
D-strings, or D5-branes on the plane-wave background analyzed in
\cite{parvizi} or branes which are of the form of fuzzy tubes or
brane on surfaces of higher genus topologies, such as the ones
constructed in \cite{BKL}.

Finally all the configurations we discussed at the level of the
TGMT are ``classical'' ones, in the sense that the entries of our
$J\by J$ matrices are $c$-numbers, rather than operators. In order
to see quantization of ${\bf J}_{12}, {\bf J}_{34}$ and
$\R_{ijab}$, which is required for matching to the $\N=4$ SYM
analysis, one needs to perform quantization. This requires a the
systematic analysis of the Tiny Graviton Matrix Perturbation
Theory. This work, which parallels a similar analysis on the BMN
matrix model  \cite{DSV1}, was started in \cite{TGMT} and is in
need of a thorough study \cite{TGMPT}.

\section*{Acknowledgments}
We would like to thank A.E. Mosaffa for fruitful discussions on
the material appeared in section 3.

\appendix
\section{Conventions and Useful Identities}
The four brackets are not as familiar objects as the usual
commutators, and in fact it is much more involved to work with
them. In this appendix we gather some useful identities for
handling and carrying out four bracket manipulations.

Let us first start with the definition of the four bracket:%
\be\begin{split}%
[A,B,C,D]  \equiv \frac{1}{24}&\Big([A,B][C,D] + [C,D][A,B]\cr &
-[A,C][B,D] - [B,D][A,C] \cr &+[A,D][B,C] + [B,C][A,D]\Big).
\end{split}\ee%
{}From the above definition one can work out the generalized
``Jacobi'' identity:%
\be\begin{split}%
[A,B,C,[D,E]] &= [[A,B,C,D],E] + [D,[A,B,C,E]] \cr &-
[[A,B],C,D,E] - [[C,A],B,D,E] - [[B,C],A,D,E].
\end{split}\ee%
As a useful special case when $E=C$, we obtain%
\be%
 [[A,B,C,D],D] =[[A,D],B,C,D] +  [A,[B,D],C,D]+[A,B,[C,D], D].%
\ee%
Trace of a four bracket is zero, \be \Tr([A,B,C,D])=0 \ee and
under the trace we have the ``by-part integration'' property, that
is \be \Tr([A,B,C,D]E)=-\Tr([A,B,C,E]D). \ee

 One can find,
through representation theory of $SO(4)$ (for
details see \cite{half-BPS}), four $\J^i$ matrices such that%
\be%
[\J^i,\J^j,\J^k,{\cal L}_5] = -\epsilon^{ijkl}\J^l
\ee%
$\J^i$ and ${\cal L}_5$ have the following explicit matrix form:%
\be%
\J^i = \left(\begin{array}{ccc} 0 & | & \Sigma^i \\ -- & - & --  \\
\bar\Sigma^i & | & 0 \end{array}\right)\ , \qquad%
{\cal L}_5= \left(\begin{array}{ccc} {\bf 1} & | & 0 \\ -- & - & --  \\
0 & | & {\bf -1} \end{array}\right)\ ,
\ee%
where $\bar\Sigma^i=(\Sigma^i)^\dagger$ and the blocks are $J/2\by
J/2$ matrices. These $\J^i$'s and $\Sigma^i$'s are generalizations
of the usual Dirac gamma matrices and the $\sigma^i$ matrices,
respectively. ${\cal L}_5$ is a direct generalization of the
chirality matrix $\gamma^5$. It is evident from the above form
that ${\cal L}_5$ {\it
anti-commutes} with $\J^i$'s, \ie%
\be\label{J-L5-anticommute}%
\{\J^i,{\cal L}_5\} = 0 %
\ee%
Therefore,%
\[%
[\J^i,\J^j{\cal L}_5]=\{\J^i,\J^j\}{\cal L}_5,\qquad [\J^i,{\cal L}_5]=-2{\cal L}_5J^i,%
\]%
moreover, ${\cal L}_5$ commutes (anti-commutes) with any product
of even (odd) number of $\J^i$'s.

Using $\J^i$ one can obtain a  $J\by J$ representation of the
$so(4)$ algebra whose generators are  $\J^{ij}\equiv [\J^i,\J^j]$.
In the explicit matrix form
\be%
\J^{ij} = \left(\begin{array}{ccc}  \Sigma^{ij}& |& 0 \\ -- & - & --  \\
0& |& \bar\Sigma^{ij}  \end{array}\right)%
\ee%
where $\Sigma^{ij}=\Sigma^i\bar\Sigma^j-\Sigma^j\bar\Sigma^i$ and
$\Sigma^{ij}=\Sigma^i\bar\Sigma^j-\Sigma^j\bar\Sigma^i$. It is
obvious that $[\J^{ij},{\cal L}_5]=0$. The $\Sigma^i$'s are chosen
in such a way that
\be [\J^{ij},\J^k] = 4i(\delta^{jk}\J^i - \delta^{ik}\J^j) \ee%

Using the above identities, and with some (perhaps tedious)
algebra one can show that
 \bse\begin{align} [\J^i,\J^j,i{\cal L}_5 \J^k,{\cal L}_5] &=
-\epsilon^{ijkl}i{\cal L}_5 \J^l +
\frac{i}{3!}(\J^j\J^kJ^i - \J^i\J^k\J^j) \\
[\J^i,i{\cal L}_5 \J^j,i{\cal L}_5 \J^k,{\cal L}_5] &=
-\epsilon^{ijkl} \J^l +
\frac{1}{3!}{\cal L}_5(\J^k\J^i\J^j - \J^j\J^i\J^k) \\
[i{\cal L}_5 \J^i,i{\cal L}_5 \J^j,i{\cal L}_5 \J^k,{\cal L}_5] &=
-\epsilon^{ijkl} i{\cal L}_5 \J^l\ .
\end{align}\ese

For studying he 1/4 BPS configurations it is useful to introduce
$\Z_0, \W_0$ matrices%
\be\begin{split} \Z_0 =& \frac{1}{\sqrt 2}(\J^1 + i\J^2) \cr \W_0
=&
\frac{1}{\sqrt 2}(\J^3 + i\J^4)\end{split}\ee%
In terms of these matrices the $SU(2)\times U(1)$ part of $SO(4)$
is manifest. (In the sense that the $U(1)$ is rotating both $\Z_0,
\W_0$ with the same phase, while under the $SU(2)$, $\left(\Z_0,\
\W_0\right)$ rotate as a doublet.) The $\J^{12},\ \J^{34}$
generators then become%
\[ \J^{12} = i[\Z_0,\bar\Z_0], \qquad \J^{34} = i[\W_0,\bar\W_0]. \]
and hence \be [[\Z_0,\bar\Z_0],\Z_0] = 4\Z_0 \ , \qquad
[[\Z_0,\bar\Z_0],\W_0]=0
\ee%

In terms of $\Z_0, \W_0$,%
\bse\begin{align}%
[\Z_0,\W_0,\bar\W_0,{\cal L}_5] =& \Z_0 \\ %
{[\bar\Z_0,\bar\W_0,\W_0,{\cal L}_5]} =& \bar\Z_0 \\ %
{[\W_0,\Z_0,\bar\Z_0,{\cal L}_5]} =& \W_0 \\ %
{[\bar\W_0,\bar\Z_0,\Z_0,{\cal L}_5]} =& \bar\W_0
\end{align}\ese%
\be{[i{\cal L}_5 \Z_0,i{\cal L}_5\W_0,i{\cal L}_5 \bar \W_0,{\cal
L}_5]} = i{\cal L}_5 \Z_0 \ee
and%
\bse\begin{align}%
[\Z_0,\W_0,i{\cal L}_5\bar \W_0,{\cal L}_5] &= [\W_0,i{\cal
L}_5\bar \W_0,\Z_0,{\cal L}_5] =
i{\cal L}_5 \Z_0 - \frac{i}{6}[\Z_0,\{\W_0,\bar \W_0\}] \\ %
{[\Z_0,i{\cal L}_5 \W_0,\bar \W_0,{\cal L}_5]} &= [i{\cal L}_5
\W_0,\bar \W_0,\Z_0,{\cal L}_5] =
i{\cal L}_5 \Z_0 + \frac{i}{6}[\Z_0,\{\W_0,\bar \W_0\}] \\ %
{[i{\cal L}_5 \Z_0,\W_0,\bar \W_0,{\cal L}_5]} &= [\W_0,\bar
\W_0,i{\cal L}_5 \Z_0,{\cal L}_5] = i{\cal L}_5 \Z_0 -
\frac{i}{3}(\W_0\Z_0\bar \W_0 - \bar \W_0\Z_0\W_0)
\end{align}\ese%
\bse\begin{align}%
[i{\cal L}_5 \Z_0,i{\cal L}_5 \W_0,\bar \W_0,{\cal L}_5] &=
[i{\cal L}_5 \W_0,\bar \W_0,i{\cal L}_5
\Z_0,{\cal L}_5] = \Z_0 - \frac{1}{6}{\cal L}_5[\Z_0,\{\W_0,\bar \W_0\}] \\
{}[i{\cal L}_5 \Z_0,\W_0,i{\cal L}_5\bar \W_0,{\cal L}_5] &=
[\W_0,i{\cal L}_5\bar \W_0,i{\cal L}_5
\Z_0,{\cal L}_5] = \Z_0 + \frac{1}{6}{\cal L}_5[\Z_0,\{\W_0,\bar \W_0\}] \\
{}[\Z_0,i{\cal L}_5 \W_0,i{\cal L}_5\bar \W_0,{\cal L}_5] &=
[i{\cal L}_5 \W_0,i{\cal L}_5\bar \W_0,\Z_0,{\cal L}_5] = \Z_0 -
\frac{1}{3}{\cal L}_5(\W_0\Z_0\bar \W_0 - \bar \W_0\Z_0\W_0)
\end{align}\ese%

\section{SUSY Generators in Terms of Matrices}\label{AppendixB}
Here we present the generators of the TGMT superalgebra \super\ in
the Matrix realization.%
\bea P^+=-P_- = \frac{1}{R_-} \Tr {\bf 1} \qquad&,&\qquad
P^-=-P_+=-{\bf H}\eea%
\be\label{Jij}
 {\bf J}_{ij}= \Tr
\big(X^{i}\Pi^{j}-X^{j}\Pi^{i}-2\theta^{\dagger\alpha\beta}
(i\sigma^{ij})_{\alpha}^{\rho}\theta_{\rho\beta} +
2\theta^{\dagger\dot{\alpha}\dot{\beta}}
(i\sigma^{ij})_{\dot{\alpha}}^{\dot{\rho}}\theta_{\dot{\rho}\dot{\beta}}\big)
\ee%
\be\label{Jab}
 {\bf J}_{ab}= \Tr
\big(X^{a}\Pi^{b}-X^{b}\Pi^{a}-2\theta^{\dagger\alpha\beta}
(i\sigma^{ab})_{\beta}^{\rho}\theta_{\alpha\rho} +
2\theta^{\dagger\dot{\alpha}\dot{\beta}}
(i\sigma^{ab})_{\dot{\beta}}^{\dot{\rho}}\theta_{\dot{\alpha}\dot{\rho}}\big)
\ee%
\be\label{Hamiltonian}
\begin{split}
{\bf H} = R_-\ \Tr&\biggl[ \frac{1}{2}(P_i^2+P_a^2) +
\frac{1}{2}\left(\frac{\mu}{R_-}\right)^2(X_i^2+X_a^2) \cr &+
\frac{1}{2\cdot 3!g_s^2} \left([ X^i , X^j , X^k, {\cal L}_5][ X^i
, X^j , X^k, {\cal L}_5] + [ X^a , X^b , X^c, {\cal L}_5][ X^a ,
X^b , X^c, {\cal L}_5]\right) \cr &+ \frac{1}{2\cdot 2g_s^2}
\left([ X^i , X^j , X^a, {\cal L}_5][ X^i , X^j , X^a, {\cal L}_5]
+ [ X^a , X^b , X^i, {\cal L}_5][ X^a , X^b , X^i, {\cal
L}_5]\right) \cr & -\frac{\mu}{3!R_- g_s}\left( \epsilon^{i j k l}
X^i [X^j, X^k, X^l, {\cal L}_5]+ \epsilon^{a b c d} X^a [ X^b,
X^c, X^d , {\cal L}_5] \right)\cr &+\left(\frac{\mu}{R_-}\right)
\left(\theta^\dagger {}^{\alpha \beta} \theta_{\alpha \beta}-
\theta_{\dot\alpha \dot\beta}\theta^\dagger {}^{\dot\alpha
\dot\beta}\right)\cr &+\frac{1}{2g_s}\left( \theta^\dagger
{}^{\alpha \beta} (\sigma^{ij})_\alpha^{\:  \: \delta} [ X^i, X^j,
\theta_{\delta \beta}, {\cal L}_5] + \theta^\dagger {}^{\alpha
\beta} (\sigma^{ab})_\alpha^{ \: \: \delta} \: [ X^a, X^b,
\theta_{\delta \beta}, {\cal L}_5]\right) \cr &+ \frac{1}{2g_s}
\left(\theta^\dagger {}^{\dot\alpha \dot\beta}
(\sigma^{ij})_{\dot\alpha}^{ \: \: \dot\delta} \: [ X^i, X^j,
\theta_{\dot\delta \dot\beta}, {\cal L}_5]+ \theta^\dagger
{}^{\dot\alpha \dot\beta} (\sigma^{ab})_{\dot\alpha}^{\: \:
\dot\delta} \: [ X^a, X^b, \theta_{\dot\delta \dot\beta},
{\cal L}_5]\right)\biggr]\,\end{split}\ee%

\be\label{Q-d-und}
\begin{split}
Q_{\dot{\alpha}\beta} = \sqrt{\frac{R_{-}}{2}}\
\Tr&\bigg[(\Pi^{i}-\frac{i\mu}{R_{-}}X^{i})
(\sigma^{i})_{\dot{\alpha}}^{\rho}\theta_{\rho\beta} +
(\Pi^{a}-\frac{i\mu}{R_{-}}X^{a})(\sigma^{a})_{\beta}^{\dot{\rho}}
\theta_{\dot{\alpha}\dot{\rho}} \cr -&
\frac{i}{3!g_{s}}\big(\epsilon^{ijkl}[X^{i},X^{j},X^{k},{\cal
L}_5] (\sigma^{l})_{\dot{\alpha}}^{\rho}\theta_{\rho\beta} +
\epsilon^{abcd}[X^{a},X^{b},X^{c},{\cal
L}_5](\sigma^{d})_{\beta}^{\dot{\rho}}
\theta_{\dot{\alpha}\dot{\rho}}\big) \cr +&
\frac{1}{2g_{s}}\big([X^{i},X^{a},X^{b},{\cal L}_5]
(\sigma^{i})_{\dot{\alpha}}^{\rho}(i\sigma^{ab})_{\beta}^{\gamma}\theta_{\rho\gamma}
+ [X^{a},X^{i},X^{j},{\cal
L}_5](\sigma^{a})_{\beta}^{\dot{\gamma}}
(i\sigma^{ij})_{\dot{\alpha}}^{\dot{\rho}}\theta_{\dot{\rho}\dot{\gamma}}\big)\bigg]
\end{split}
\ee%
\be\label{Q-und-d}
\begin{split} Q_{\alpha\dot\beta} = \sqrt{\frac{R_{-}}{2}}\
\Tr&\bigg[(\Pi^{i}-\frac{i\mu}{R_{-}}X^{i})
(\sigma^{i})_{\alpha}^{\dot\rho}\theta_{\dot\rho\dot\beta} +
(\Pi^{a}-\frac{i\mu}{R_{-}}X^{a})(\sigma^{a})_{\dot\beta}^{\rho}
\theta_{\alpha\rho} \cr -&
\frac{i}{3!g_{s}}\big(\epsilon^{ijkl}[X^{i},X^{j},X^{k},{\cal
L}_5] (\sigma^{l})_{\alpha}^{\dot\rho}\theta_{\dot\rho\dot\beta} +
\epsilon^{abcd}[X^{a},X^{b},X^{c},{\cal
L}_5](\sigma^{d})_{\dot\beta}^{\rho} \theta_{\alpha\rho}\big) \cr
+& \frac{1}{2g_{s}}\big([X^{i},X^{a},X^{b},{\cal L}_5]
(\sigma^{i})_{\alpha}^{\dot\rho}(i\sigma^{ab})_{\dot\beta}^{\dot\gamma}
\theta_{\dot\rho\dot\gamma} + [X^{a},X^{i},X^{j},{\cal
L}_5](\sigma^{a})_{\dot\beta}^{\gamma}
(i\sigma^{ij})_{\alpha}^{\rho}\theta_{\rho\gamma}\big)\bigg]
\end{split}\ee%
\be\label{R} \R^{ijab} =
\frac{1}{g_s}\Tr\big([X^{i},X^{j},X^{a},X^{b}]{\cal L}_5\big) \ee%
\be\begin{split} C^{ia} = \frac{R_-}{\mu}\Tr &\bigg[P^iP^a -
\left(\frac{1}{2g_s}\right)^2
\epsilon^{abcd}\epsilon^{ijkl}[X^{j},X^{b},X^{c},{\cal L}_5]
[X^{d},X^{k},X^{l},{\cal L}_5] \cr  +&
\Big(\frac{\mu}{R_{-}}X^{i}+\frac{1}{3!g_s}\epsilon^{ijkl}[X^{j},X^{k},X^{l},{\cal L}_5]\Big)\Big(\frac{\mu}{R_{-}}X^{a}+\frac{1}{3!g_s}\epsilon^{abcd}[X^{b},X^{c},X^{d},{\cal L}_5]\Big)\bigg]\end{split}\ee%
\be \hat C^{ia} =
\frac{R_-}{\mu}\frac{1}{2g_s}\Tr\Big(\epsilon^{ijkl}P^j[X^a,X^k,X^l,{\cal
L}_5]
+ \epsilon^{abcd}P^c[X^i,X^c,X^d,{\cal L}_5]\Big) \ee%
%
\end{document}